\documentclass[12pt]{article}
\usepackage{latexsym,amssymb}
\usepackage{amsmath,amsbsy}
\usepackage{amsopn,amstext}
\usepackage{cite}
\usepackage{amssymb}
\usepackage{rotating}
\usepackage{amsmath}
\usepackage{color}
\usepackage{makecell}
\usepackage{multirow}

\allowdisplaybreaks[4]

\renewcommand{\theequation}{\arabic{section}.\arabic{equation}}

\begin{document}



\def\a{\alpha}
\def\b{\beta}
\def\d{\delta}
\def\e{\epsilon}
\def\g{\gamma}
\def\h{\mathfrak{h}}
\def\k{\kappa}
\def\l{\lambda}
\def\o{\omega}
\def\p{\wp}
\def\r{\rho}
\def\t{\tau}
\def\s{\sigma}
\def\z{\zeta}
\def\x{\xi}
\def\V={{{\bf\rm{V}}}}
 \def\A{{\cal{A}}}
 \def\B{{\cal{B}}}
 \def\C{{\cal{C}}}
 \def\D{{\cal{D}}}
\def\G{\Gamma}
\def\K{{\cal{K}}}
\def\O{\Omega}
\def\R{\bar{R}}
\def\T{{\cal{T}}}
\def\L{\Lambda}
\def\f{E_{\tau,\eta}(sl_2)}
\def\E{E_{\tau,\eta}(sl_n)}
\def\Zb{\mathbb{Z}}
\def\Cb{\mathbb{C}}

\def\R{\overline{R}}

\def\beq{\begin{equation}}
\def\eeq{\end{equation}}
\def\bea{\begin{eqnarray}}
\def\eea{\end{eqnarray}}
\def\ba{\begin{array}}
\def\ea{\end{array}}
\def\no{\nonumber}
\def\le{\langle}
\def\re{\rangle}
\def\lt{\left}
\def\rt{\right}

\baselineskip=20pt

\newfont{\elevenmib}{cmmib10 scaled\magstep1}
\newcommand{\preprint}{
   \begin{flushleft}
   \end{flushleft}\vspace{-1.3cm}
   \begin{flushright}\normalsize
   \end{flushright}}
\newcommand{\Title}[1]{{\baselineskip=26pt
   \begin{center} \Large \bf #1 \\ \ \\ \end{center}}}

\newcommand{\Author}{\begin{center}
   \large \bf
Guang-Liang Li${}^{a,b}$, Junpeng
Cao${}^{b,c,d,e,}\footnote{Corresponding author}$, Pei
Sun${}^{b,f,g}$, Wen-Li
Yang${}^{b,f,g,1}$, Kangjie Shi${}^{f}$ and Yupeng Wang${}^{c}$
 \end{center}}

\newcommand{\Address}{\begin{center}
${}^a$ Ministry of Education Key Laboratory for Nonequilibrium
Synthesis and Modulation of Condensed Matter, School of
Physics, Xi'an Jiaotong University, Xi'an 710049, China\\
${}^b$ Peng Huanwu Center for Fundamental Theory, Xi'an 710127, China\\
${}^c$ Beijing National Laboratory for Condensed Matter Physics, Institute of Physics, Chinese Academy of Sciences, Beijing 100190, China\\
${}^d$ School of Physical Sciences, University of Chinese Academy of Sciences, Beijing 100049, China\\
${}^e$ Songshan Lake Materials Laboratory, Dongguan, Guangdong 523808, China \\
${}^f$ Institute of Modern Physics, Northwest University, Xi'an 710127, China\\
${}^g$ Shaanxi Key Laboratory for Theoretical Physics Frontiers, Xi'an 710127, China\\
E-mail: leegl@xjtu.edu.cn, junpengcao@iphy.ac.cn, wlyang@nwu.edu.cn, kjshi@nwu.edu.cn,
yupeng@iphy.ac.cn
\end{center}}

\preprint
\thispagestyle{empty}
\bigskip\bigskip\bigskip

\Title{Exact solution of a quantum integrable system associated with the $G_2$ exceptional Lie algebra} \Author

\Address
\vspace{1cm}

\begin{abstract}

A quantum integrable spin chain model  associated with the $G_2$ exceptional Lie algebra is studied.
By using the fusion technique,  the closed recursive relations among the fused transfer matrices are obtained.
These identities allow us to derive the exact energy spectrum and  Bethe ansatz equations of the system based on polynomial analysis.
The present method provides a unified treatment to investigate the Bethe ansatz solutions for both the periodic and  the non-diagonal open boundary conditions associated with exceptional Lie algebras.

\vspace{1truecm} \noindent {\it PACS:} 75.10.Pq, 02.30.Ik, 71.10.Pm

\noindent {\it Keywords}: Bethe Ansatz; Lattice Integrable Models
\end{abstract}
\newpage

\section{Introduction}

Quantum integrable systems possess rich applications in several important research fields  such as statistical mechanics, condensed matter physics, theoretical and mathematical physics \cite{1,2,3}.
Symmetries  play a crucial role in physical systems and  quantum integrable models with different symmetries may clarify different universal class in physics world \cite{4}. The quantum integrable systems with symmetries characterized by
the $A$ \cite{A-1,A-2,A-3} and $B,C, D$ \cite{B-1,B-2,B-3,B-4,B-5,C-1,C-2,C-3} series Lie algebras have
been studied extensively and  the exact Bethe ansatz solutions of a vast amount of models have been obtained for both periodic and open
boundaries \cite{5}.

We note that there do exist some physical systems associated with exception Lie algebras such as $G_2$ and $E_8$.
These symmetries can induce interesting physics and have potential impacts on topological phases and topological quantum computation \cite{i1, i2, i3, Li23, Hu18, Lop19, Nay08}. For example, the Ising model with $E_8$ symmetry has several different ordered states \cite{i1,i2,i3}.
The $G_2$ symmetry is the smallest possible exceptional Lie algebra besides the automorphism group of the algebra of octonions \cite{john1}.
Its relation to Clifford algebras and spinors, Bott periodicity, projective
and Lorentzian geometry, Jordan algebras, and the exceptional Lie groups has been studied.
The holonomy group $G_2$ is also associated with
the compact Riemannian manifolds with special geometric structure, such as Spin-7 manifolds or nearly
K\"{a}hler manifolds \cite{john2}. These manifolds
play important roles as ingredients for compactifications in string theory, topology and
M-theory. In addition, the $G_2$ model has applications in quantum logic, special relativity and
supersymmetry \cite{john3}.
The Bethe ansatz solutions of the $G_2$ model related to the $R$-matrix associated with the $G_2$ exceptional Lie algebra \cite{Ogi86, Mac91, martins} with periodic boundary condition was studied by Martins \cite{martins}, and the diagonal open boundary condition was studied by Yung and Batchelor \cite{yung}.

In this paper, we study the exact solution of the $G_2$ model with
non-diagonal open boundary condition. By solve the reflection
equations \cite{re-1,re-2}, we obtain the reflection matrices with
non-diagonal elements, which indicates the $U(1)$ symmetry  in the bulk  is broken \cite{3,a1,a2} and the conventional Bethe ansatz method can not be used to approach this model. Alternatively, we adopt
 the fusion techniques \cite{f1,f2,f3,f4,f5,f6,f7} and  the off-diagonal Bethe ansatz method developed in \cite{Cao1,Cao13,Cao14,Hao14,Cao2} to derive the exact spectrum of the model.

The paper is organized as follows. In section 2, we introduce the $R$-matrix and its properties of the $G_2$ model.
In section 3, we give the solution of reflection equations. Based on them, we construct the transfer matrix and the model Hamiltonian.
In section 4, we construct a closed set of operator identies of the  transfer matrices as well as  the asymptotic behaviors and the values at some special points of the transfer matrices.
In section 5, we list all the necessary functional relations to determine the
eigenvalues of the transfer matrices, which allow us to obtain the eigenvalues and express them in terms of the inhomogeneous $T-Q$ relations.
The related Bethe ansatz equations are also given.
In section 6, we study the exact solution of the system with periodic boundary condition.
Concluding remarks are given in section 7. Appendices A-D supply  some technical derivations.

\section{$R$-matrix and its properties of the $G_2$ model}
\setcounter{equation}{0}

Let ${\rm\bf V}$ denote a $7$-dimensional linear space with an orthonormal basis $\{|j\rangle, j=1,\cdots, 7\}$, which endows the $7$-dimensional
representation of the exceptional $G_2$ Lie algebra. In this paper, we always adopt the convention: For a matrix $A\in {\rm End}({\rm\bf V})$, $A_j$ is an
embedding operator in the  tensor space
${{\rm\bf V}}\otimes {{\rm\bf V}}\otimes\cdots$, which acts as $A$ on the $j$-th
space and as identity on the other factor spaces. For the matrix $R\in {\rm End}({ {\rm\bf V}}\otimes { {\rm\bf V}})$, $R_{ij}$ is an embedding
operator in the  tensor space, which acts as identity on the factor spaces except for the $i$-th and $j$-th ones.

The quantum integrable model associated with the $G_2$ exceptional Lie algebra is quantified by the
$49\times49$ $R$-matrix defined in the ${\bf V}\otimes {\bf V}$
space  \cite{Ogi86, Mac91, martins}
\begin{eqnarray}
   &&\hspace{-0.8truecm}R_{12}(u) =a(u)\sum_{i=1,i\neq 4}^{7}(E^{i}_i\otimes E_i^i)+\bar{a}(u)(E^{4}_4\otimes E_4^4)
   +c(u)\sum_{i=1,i\neq 4}^{7}(E^{i}_i\otimes E_4^4+E^{4}_4\otimes E_i^i)\no\\
   &&+e(u)\sum_{i=1}^{3}(E^{i}_i\otimes E^{\bar{i}}_{\bar{i}}+ E^{\bar{i}}_{\bar{i}}\otimes E_i^i)
   +b(u)\sum_{i=2}^{3}(E^1_1\otimes E^{i}_{i}+ E^{i}_{i}\otimes E_1^1+E^7_7\otimes E^{\bar{i}}_{\bar{i}}+ E^{\bar{i}}_{\bar{i}}\otimes E_7^7\no\\
   &&+E^i_i\otimes E^{i+3}_{i+3}+ E^{i+3}_{i+3}\otimes E_i^i)+d(u)\sum_{i=5}^{6}(E^1_1\otimes E^{i}_{i}+ E^{i}_{i}\otimes E_1^1+E^7_7\otimes E^{\bar{i}}_{\bar{i}}+ E^{\bar{i}}_{\bar{i}}\otimes E_7^7\no\\
   &&+E^i_i\otimes E^{\bar{i}+3}_{\bar{i}+3}+E^{\bar{i}}_{\bar{i}}\otimes E_{i-3}^{i-3})
   +g_1(u)\sum_{i=2}^{3}(E^1_i\otimes E^{i}_{1}+ E^{i}_{1}\otimes E_i^1+E^7_{\bar{i}}\otimes E^{\bar{i}}_{7}+ E^{\bar{i}}_{7}\otimes E_{\bar{i}}^7\no\\
   &&+E^i_{i+3}\otimes E^{{i}+3}_{i}+E^{{i}+3}_{i}\otimes E_{i+3}^i)
   +g_6(u)\sum_{i=5}^{6}(E^1_i\otimes E^{i}_{1}+ E^{i}_{1}\otimes E_i^1+E^7_{\bar{i}}\otimes E^{\bar{i}}_{7}+ E^{\bar{i}}_{7}\otimes E_{\bar{i}}^7\no\\
   &&+E^i_{\bar{i}+3}\otimes E^{\bar{i}+3}_{i}+E^{\bar{i}}_{i-3}\otimes E_{\bar{i}}^{i-3})
   +g_4(u)\sum_{i=1,i\neq 4}^{7}(E^4_i\otimes E^{i}_{4}+ E^{i}_{4}\otimes E_i^4)\no\\
   &&+g_8(u)\sum_{i=1}^3(E^i_{\bar{i}}\otimes E^{\bar{i}}_{i}+ E^{\bar{i}}_{i}\otimes E_{\bar{i}}^i)+g_5(u)\sum_{i=1}^3\xi_i(E^i_{4}\otimes
    E^{\bar{i}}_{4}+E^{\bar{i}}_{4}\otimes E_{4}^{i}+E_i^{4}\otimes E_{\bar{i}}^{4}+E_{\bar{i}}^{4}\otimes E^{4}_{i})\no\\
    &&+g_3(u)\sum_{i=2}^3[E^i_{i+3}\otimes E^{\bar{i}}_{\bar{i}-3}+E^{\bar{i}}_{\bar{i}-3}\otimes E^i_{i+3}-\xi_i(E^1_i\otimes E^7_{\bar{i}}+ E^7_{\bar{i}}\otimes E^1_i+E^i_1\otimes E^{\bar{i}}_7+ E^{\bar{i}}_7\otimes E^i_1)
    ]\no\\
    &&+g_7(u)\sum_{i=2}^3[E^i_{\bar{i}-3}\otimes E^{\bar{i}}_{i+3}+E^{\bar{i}}_{i+3}\otimes E^i_{\bar{i}-3}-\xi_i(E^1_{\bar{i}}\otimes E^7_{i}+ E^7_i\otimes E^1_{\bar{i}}+E^{\bar{i}}_1\otimes E^{i}_7+ E^{i}_7\otimes E^{\bar{i}}_1)
    ]\no\\
    &&+g_2(u)\sum_{i\le j,l\le k,i\neq \bar{l},j\neq \bar{k},i+k=j+l=5,6,7,9,10,11}(E^i_j\otimes E^k_l+E^j_i\otimes E^l_k-E^i_l\otimes E^k_j-E^j_k\otimes
    E^l_i\no\\[6pt]
    &&\hspace{20mm}+E^k_l\otimes E^i_j+E^l_k\otimes E^i_j-E^k_j\otimes E^i_l-E^l_i\otimes
    E^j_k ),
 \label{rm}
\end{eqnarray}
where $u$ is the spectral parameter, $\{E_i^j|i, j =1, \cdots, 7\} $ are the Weyl basis and the non-vanishing matrix elements are
\begin{eqnarray}
&&a(u)=(u+1)(u+4)(u+6),\quad \bar{a}(u)=(u+2)(u+3)(u+4),\quad
b(u)=u(u+4)(u+6),\no\\
&& c(u)=u(u+3)(u+6),\quad d(u)=u(u+2)(u+6),\quad
e(u)=u(u+2)(u+5),\no\\
&& g_1(u)=(u+4)(u+6),\quad g_2(u)=\sqrt{2}u(u+6),\quad
g_3(u)=u(u+2),\no\\
&&g_4(u)=2(u+2)(u+6),\quad g_5(u)=2 u(u+4),\quad
g_6(u)=(3u+4)(u+6),\no\\
&& g_7(u)=u(3u+14),\quad g_8(u)=8(u+3).\no
\end{eqnarray}
Here and after we use the convention:  $i+\bar{i}=8$ and $\xi_1=-\xi_2=\xi_3=1$.
The $R$-matrix satisfies the quantum Yang-Baxter equation
\begin{eqnarray}
R_{12}(u-v)R_{13}(u)R_{23}(v)=R_{23}(v)R_{13}(u)R_{12}(u-v).
\label{YBE}
\end{eqnarray}
Moreover, it also satisfies the very properties \cite{martins}
\begin{eqnarray}
\hspace{-0.8truecm}{\rm regularity:}&&R_{12}(0)=\rho_{12}(0)^{\frac{1}{2}}{\cal P}_{12},\\[4pt]
\hspace{-0.8truecm}{\rm unitary:}&&R_{12}(u)R_{21}(-u)=a_1(u)a_1(-u)\equiv \rho_{12}(u),\\[4pt]
\hspace{-0.8truecm}{\rm
crossing \; symmetry:}&&R_{12}(u)=-V_1R^{t_1}_{21}(-u-6)V_1^{-1}=-V_2^{t_2}R^{t_2}_{21}(-u-6)[V_2^{t_2}]^{-1},\label{C-S}
\end{eqnarray}
where ${\cal P}_{12}$ is the permutation operator with the matrix
elements $[{\cal P}_{12}]^{ij}_{kl}=\delta_{il}\delta_{jk}$, $t_i$
denotes the transposition in the $i$-th space, $R _{21}={\cal
P}_{12}R _{12}{\cal P}_{12}$, the elements of crossing matrix $V_1$ (or $V_2$) are
$V_{ij}=(-1)^{i-1}\delta_{i,\bar{j}}$ where $i$ and $j$ are the row and column indices, respectively. Combining  the
crossing symmetry and the unitary of the $R$-matrix, one can derive
\begin{eqnarray}
{\rm
crossing\; unitary:}&&R^{t_1}_{12}(u)R^{t_1}_{21}(-u-12)=-\rho_{12}(u+6)\equiv \tilde{\rho}_{12}(u),\label{Permutation-1}
\end{eqnarray}
which will be useful in the following parts of the paper.

\section{Integrable open $G_2$ chain}
\setcounter{equation}{0}

Now, we construct the integrable $G_2$ model with open boundary condition. We first define
the single-row monodromy matrix
\bea
T_0(u)=R_{01}(u-\theta_1)R_{02}(u-\theta_{2})\cdots
R_{0N}(u-\theta_N), \label{Mon-1} \eea where the index $0$
indicates the auxiliary space ${\bf V}_0$, the other tensor space ${\bf
V}_1 \otimes \cdots \otimes {\bf V}_N$ is the quantum
space, $N$ is the number of sites and $\{\theta_j|j=1,\cdots, N\}$
are the inhomogeneous parameters. Thanks to the quantum Yang-Baxter equation (\ref{YBE}), the monodromy matrix satisfies
the Yang-Baxter relation
\bea
 R_{12}(u-v) T_1(u) T_2(v) = T_2(v) T_1(u) R_{12}(u-v).\label{YR-1}
\eea

For the integrable open chain, the boundary reflections at one end
are characterized by the reflection matrix $K^{-}(u)$. The
integrability requires that the reflection matrix satisfies the
reflection equation (RE) \cite{re-1,re-2}
\begin{equation}
 R_{12}(u-v){K^{  -}_{  1}}(u)R_{21}(u+v) {K^{   -}_{2}}(v)=
 {K^{   -}_{2}}(v)R_{12}(u+v){K^{   -}_{1}}(u)R_{21}(u-v).
 \label{RE1}
 \end{equation}
By solving Eq.\eqref{RE1}, we obtain the reflection matrix which has the non-diagonal elements
\bea  K^{ -}(u)=1+M u,\quad
 M=\left(\begin{array}{ccccccc}c_{11}&0&0&0&c_{1}&c_{2}&0\\[6pt]
0&c_{22}&c_3&0&0&0&-c_2\\[6pt]
0&c_{3}&c_{33}&0&0&0&c_1\\[6pt]
0&0&0&-2&0&0&0\\[6pt]
c_1&0&0&0&c_{33}&-c_3&0\\[6pt]
c_2&0&0&0&-c_3&c_{22}&0\\[6pt]
0&-c_2&c_1&0&0&0&c_{11}
\end{array}\right),\label{K-matrix-VV1} \eea
where $c_1, c_2, c_3$ are the boundary parameters and   \bea
c_{11}=\frac{c_1c_3}{c_2}+\frac{c_2c_3}{c_1}-2,\quad
c_{22}=2-\frac{c_2c_3}{c_1},\quad c_{33}=2-\frac{c_1c_3}{c_2}.
\no\eea Meanwhile, the boundary parameters need to satisfy the
constraint \bea
\frac{c_1c_3}{c_2}+\frac{c_2c_3}{c_1}+\frac{c_1c_2}{c_3}=4.\eea
Thus there are two free parameters. When $c_1=0, c_2=0, c_3=2$, the
reflection matrix (\ref{K-matrix-VV1}) becomes
\bea  K^{ -}_c(u)=\left(\begin{array}{ccccccc}1+2u&0&0&0&0&0&0\\[6pt]
0&1&2u&0&0&0&0\\[6pt]
0&2u&1&0&0&0&0\\[6pt]
0&0&0&1-2u&0&0&0\\[6pt]
0&0&0&0&1&-2u&0\\[6pt]
0&0&0&0&-2u&1&0\\[6pt]
0&0&0&0&0&0&1+2u
\end{array}\right).\label{K-matrix-dig1} \eea
We find that the $K^{ -}_c(u)$ matrix can be obtained by a diagonal matrix with a gauge
transformation, i.e., \bea K^{
-}_c(u)=G^{-1}K^-_d(u) G,\label{q} \eea where $K^{ -}_d(u)$ is a diagonal matrix with the form \bea
K^-_d(u)=Diag(1+2u,1+2u,1-2u,1-2u,1-2u,1+2u,1+2u),\eea
which agrees with that given in ref.[30] after taking the
rational limit. The gauge
transformation $G$ in Eq.(\ref{q}) is
\bea G=
\left(\begin{array}{ccccccc}1&1&1&0&\sqrt{2}-\frac{1}{2}&\frac{1}{2}-\sqrt{2}&1-2\sqrt{2}\\[6pt]
\sqrt{2}&1&1&0&2-\frac{1}{\sqrt{2}}&\frac{1}{\sqrt{2}}-2&1-2\sqrt{2}\\[6pt]
0&-1&1&2-\sqrt{2}&2&2&0\\[6pt]
0&-2&2&2(1+\sqrt{2})&2\sqrt{2}&2\sqrt{2}&0\\[6pt]
0&-2&2&4&2&2&0\\[6pt]
-2&-2&-2&0&-1&1&2\\[6pt]
-2\sqrt{2}&-2&-2&0&-\sqrt{2}&\sqrt{2}&2
\end{array}\right), \eea
which satisfies the relation \bea[G\otimes G, R(u)]=0.\label{q1}
\eea

Due to the reflection, we should define the reflecting single-row monodromy matrix
\begin{eqnarray}
\hat{T}_0 (u)=R_{N0}(u+\theta_N)\cdots R_{20}(u+\theta_{2}) R_{10}(u+\theta_1),\label{Tt11}
\end{eqnarray}
which characterizes the reflected quasi-particle scattering with others.
The reflecting monodromy matrix (\ref{Tt11}) satisfies the Yang-Baxter relation
\begin{eqnarray}
R_{ 21} (u-v) \hat T_{1}(u) \hat T_2(v)=\hat T_2(v) \hat T_{ 1}(u)
R_{21} (u-v).\label{YR-2}
\end{eqnarray}
From \eqref{Mon-1}, \eqref{K-matrix-VV1} and \eqref{Tt11}, we define the double-row monodromy matrix as
\begin{equation}
U(u)= T_0 (u) K^{-}_0(u)\hat{T}_0 (u).
\end{equation}
The boundary reflections at the other end of the chain is characterized by the dual reflection matrix
\begin{equation}
K^{ +}(u)=K^{-}(-u-6)|_{\{c_1,c_2,c_3\}\rightarrow
\{\tilde{c}_1,\tilde{c}_2,\tilde{c}_3\}}, \label{ksk}
\end{equation}
where $\tilde{c}_1,\tilde{c}_2$ and $\tilde{c}_3$ are the
boundary parameters. The $K^{ +}(u)$ satisfies the dual RE
\begin{eqnarray}
 &&R_{21}(u-v){K^{   +}_{2}}(v)R_{12}
 (-u-v-12){K^{   +}_{1}}(u)\nonumber\\[4pt]
&&\qquad\qquad\quad\quad={K^{   +}_{1}}(u)R_{21}(-u-v-12) {K^{
+}_{2}}(v)R_{12}(u-v).
 \label{RE2}
\end{eqnarray}
Then the complete scattering and reflection processes of the quasi-particle are characterized by the transfer matrix \cite{re-1}
\begin{eqnarray}
t(u)= tr_0 \{K_0^{ +}(u)U(u)\}. \label{t-1}
\end{eqnarray}
From the Yang-Baxter relation, RE and dual RE, one can prove that the transfer matrices with different spectral parameters commute with each other, i.e., $[t(u), t(v)]=0$.
Therefore, $t(u)$ serves as the generating function of the conserved
quantities in the system.
The model Hamiltonian can
be obtained by taking the derivative of the logarithm of the
transfer matrix as \cite{re-1}
\bea
H&=&\frac{\partial \ln t(u)}{\partial
u}|_{u=0,\{\theta_j\}=0} \nonumber \\[8pt]
&=& \sum^{N-1}_{k=1}H_{k k+1}+\frac{1}{2}{K^{-}_N}'(0)+\frac{ tr_0
\{K^{+}_0(0)H_{10}\}}{tr_0 K^{+}_0(0)}+{\rm constant}. \label{hh}
\eea

We remark that the reflection matrices $K^{\pm}(u)$ are generally
not commutative and therefore  the $U(1)$ symmetry of the system
is broken, which makes the algebraic Bethe ansatz hard to be used.

\section{Function relations of the transfer matrices}
From the definition of transfer matrix (\ref{t-1}),
we know that $t(u)$ is a operator polynomial of $u$ with the
degree $6N+2$. Then the value of $t(u)$ can be determined by the
$6N+3$ constraints satisfied by $t(u)$. For this purpose, we take the method developed in \cite{Cao1,Cao13,Cao14,Hao14,Cao2} as follows. Firstly, let us introduce some "auxiliary" commutative transfer matrices (e.g., the fused transfer matrices $\bar{t}(u)$ and $\tilde{t}(u)$ (see below (\ref{t-2}) and (\ref{t-3})) by the fusion technique \cite{f1,f2,f3,f4,f5,f6,f7}, which commute with the fundamental transfer matrix $t(u)$.
Based on the polynomial analysis, we seek for sufficient constraint conditions (see below (\ref{Transfer-Crossing}), (\ref{Op-pro-open-2}), (\ref{Op-pro-open-3}), (\ref{Crossing-2}) and (\ref{Op-pro-open-4})-(\ref{sp-5})) to determine the eigenvalues of all the transfer matrices in this section.

Based on the crossing relation of the fundamental $R$-matrix (\ref{C-S}), we can show that the transfer matrix $t(u)$ satisfies the
crossing relation
\begin{eqnarray}
t(u)=t(-u-6). \label{Transfer-Crossing}
\end{eqnarray}
The detailed proof is given in Appendix B.

The property (\ref{pro-0}) and the quantum Yang-Baxter equation (\ref{YBE}) allow us to arrive at
\bea
&&P^{(1)}_{21}R_{13}(u)R_{23}(u-6)P^{(1)}_{21}=a(u)e(u-6)\times
{\rm id},\no \\[6pt]
&&P^{(1)}_{12}R_{31}(u)R_{32}(u-6)P^{(1)}_{12}=a(u)e(u-6)\times
{\rm id}, \label{Q-det}
\eea
where the $1$-dimensional projector
$P^{(1)}_{12}=P^{(1)}_{21}$ is given by (\ref{a1})-(\ref{1-dim}).
The above relations imply that the product of the transfer
matrices satisfies the relation \bea &&
\hspace{-1.8truecm}t(\theta_j)\,t(\theta_j-6)=4^2
\frac{(\theta_j-1)(\theta_j-6)
(\theta_j+1)(\theta_j+6)}{(\theta_j-2)(\theta_j-3)
(\theta_j+2)(\theta_j+3)}\no\\[6pt]
&&\hspace{-1.2truecm}\times
(\theta_j-\frac{1}{{2}})(\theta_j-\frac{5}{{2}})
(\theta_j+\frac{1}{{2}})(\theta_j+\frac{5}{{2}})\prod_{i=1}^N{\rho}_{12}(\theta_j-\theta_i){\rho}_{12}(\theta_j+\theta_i)\times {\rm id},\; j=1,\cdots, N.
\label{Op-pro-open-2}
\eea
The detailed proof is given in Appendix C.

\subsection{Fused transfer matrix}

By using the fused $R$-matrix (\ref{1YBE15}), we can introduce the fused monodromy matrix and the reflecting one as
 \bea
&&T_{\bar{0}}(u)=R_{{\bar{0}}1}(u-\theta_1)R_{{\bar{0}}2}(u-\theta_{2})\cdots
R_{{\bar{0}}N}(u-\theta_N), \label{Mon-11}\\[6pt]
&&\hat{T}_{\bar{0}} (u)=R_{N{\bar{0}}}(u+\theta_N)\cdots
R_{2{\bar{0}}}(u+\theta_{2})
R_{1{\bar{0}}}(u+\theta_1).\label{Mon-21} \eea

By using the fused monodromy matrices (\ref{Mon-11})-(\ref{Mon-21}) and fused
reflection matrices (\ref{RE15-1})-(\ref{RE15-2}), we construct the fused transfer matrix
$\bar{t}(u)$ as
\begin{equation} \bar{t}(u)= tr_{\bar{0}} \{K_{\bar{0}}^{   +}(u)T_{\bar{0}} (u) K^{
 -}_{\bar{0}}(u)\hat{T}_{\bar{0}} (u)\}, \label{t-2}
\end{equation}
which commute with the fundamental transfer matrix $t(u)$ and also itself, i.e., $[t(u),\,\bar{t}(v)]=[\bar{t}(u),\,\bar{t}(v)]=0$.
It can be shown that the product of the fundamental transfer matrices at some special points have
the relations
\bea
&&\hspace{-2.2truecm}
t(\pm\theta_j)\,t(\pm\theta_j-1)=-
\frac{(\pm\theta_j-1)(\pm\theta_j+6)(\pm\theta_j+\frac{5}{{2}})^2
}{(\pm\theta_j+2)(\pm\theta_j+3)
}\prod_{i=1}^N[(\pm\theta_j-\theta_i-1)\no\\[6pt]
&&\hspace{1mm}\times
(\pm\theta_j+\theta_i-1)a(\pm\theta_j-\theta_i)a(\pm\theta_j+\theta_i)]\,
\bar{t}(\pm\theta_j-\frac{1}{{2}}),\; j=1,\cdots, N. \label{Op-pro-open-3}
\eea
The detailed proof is given in Appendix C. It is noted that the relations (\ref{Op-pro-open-3}) with $+\theta_j$ and that with $-\theta_j$ are indeed
independent. Moreover, we can show the fused transfer matrix $\bar{t}(u)$ also satisfies the crossing relation
\bea
\bar{t}(u)= \bar{t}(-u-6).\label{Crossing-2}
\eea
We shall shift the detailed proof of the very relation in Appendix B.

Moreover, by using the next fused $R$-matrices (\ref{oip1}), we introduce the next fused monodromy matrices as
\bea
&&T_{\tilde{0}}(u)=R_{\tilde{0}1}(u-\theta_1)R_{{\tilde{0}}2}(u-\theta_{2})\cdots
R_{{\tilde{0}}N}(u-\theta_N), \label{Mon-5}\\[6pt]
&&\hat{T}_{\tilde{0}} (u)=R_{N{\tilde{0}}}(u+\theta_N)\cdots
R_{2{\tilde{0}}}(u+\theta_{2})
R_{1{\tilde{0}}}(u+\theta_1),\label{Mon-6} \eea
and the associated fused transfer matrix $\tilde{t}(u)$
\begin{equation} \tilde{t}(u)= tr_{\tilde{0}} \{K_{\tilde{0}}^{   +}(u)T_{\tilde{0}} (u) K^{
 -}_{\tilde{0}}(u)\hat{T}_{\tilde{0}} (u)\}. \label{t-3}
\end{equation}
With the help of the fusion procedure, it is easy to check that the transfer matrices $t(u)$, $\bar{t}(u)$ and $\tilde{t}(u)$ commute with each other,
\bea
[t(u),\,t(v)]=[t(u),\,\bar{t}(v)]=[t(u),\,\tilde{t}(v)]=[\bar{t}(u),\,\bar{t}(v)]=[\bar{t}(u),\,\tilde{t}(v)]=[\tilde{t}(u),\,\tilde{t}(v)]=0.\no
\eea
We can show that the resulting transfer matrices satisfy the product relations and the crossing relation
\bea
&&\hspace{-2.2truecm} t(\pm\theta_j)\,\bar{t}(\pm\theta_j-\frac{7}{{2}})=-
\frac{(\pm\theta_j+1)(\pm\theta_j+6) }{(\pm\theta_j-\frac{1}{{2}})
(\pm\theta_j-\frac{3}{{2}})(\pm\theta_j+3)(\pm\theta_j+4)}\no\\[6pt]
&&\hspace{5mm}\times
\prod_{i=1}^N(\pm\theta_j-\theta_i+6)(\pm\theta_j+\theta_i+6)\,
\tilde{t}(\pm\theta_j-\frac52), \; j=1,\cdots, N, \label{Op-pro-open-4}\\[6pt]
&&\hspace{-2.2truecm}\tilde{t}(u)=\tilde{t}(-u-6). \label{Crossing-3}
 \eea
Moreover, we can further prove that
\bea
&&\hspace{-1.2truecm}
t(\pm\theta_j)\,\tilde{t}(\pm\theta_j-\frac92)=2^4
\frac{(\pm\theta_j+1)(\pm\theta_j+6) }{(\pm\theta_j+3)
(\pm\theta_j+4)}(\pm\theta_j-\frac12)(\pm\theta_j-\frac{7}{2})(\pm\theta_j+\frac12)(\pm\theta_j+\frac{5}{2})\no\\[6pt]
&&\hspace{5mm}\times
\prod_{i=1}^N(\pm\theta_j-\theta_i+6)(\pm\theta_j+\theta_i+6)
t(\pm\theta_j-2)t(\pm\theta_j-5), \; j=1,\cdots, N, \label{Op-pro-open-5}\\[6pt]
&&\hspace{-1.2truecm}
t(\pm\theta_j)\,\tilde{t}(\pm\theta_j-\frac{13}{2})=-2^6
\frac{(\pm\theta_j-4)(\pm\theta_j+1)(\pm\theta_j+6)
}{(\pm\theta_j-2)(\pm\theta_j+2)
(\pm\theta_j+3)}(\pm\theta_j-\frac{11}{2})(\pm\theta_j-\frac52)\no\\[4pt]
&&\hspace{5mm}\times(\pm\theta_j-\frac{3}{2})(\pm\theta_j-\frac{1}{2})(\pm\theta_j+\frac12)
(\pm\theta_j+\frac{5}{2})\prod_{i=1}^N[(\pm\theta_j-\theta_i+4)\no\\[4pt]
&&\hspace{5mm}\times
(\pm\theta_j+\theta_i+4)a(\pm\theta_j-\theta_i)a(\pm\theta_j+\theta_i)]\,
t(\pm\theta_j-7),\; j=1,\cdots, N. \label{Op-pro-open-6}
\eea
The detailed proofs of the relations (\ref{Op-pro-open-4})-(\ref{Op-pro-open-6}) are given in Appendices B\&C with the help of the
fusion properties of the $R$-matrices obtained in Appendix A.

\subsection{Asymptotic behaviors and special points}

Taking the limit of $u$ tends to infinity and using the definitions, we obtain the asymptotic behaviors of the transfer matrices as
\bea && t(u)|_{u\rightarrow \pm\infty}= A u^{6N+2}\times {\rm id}+\cdots, \label{as-1} \\[4pt]
&& \bar{t}(u)|_{u\rightarrow \pm\infty}=  16[-(\frac{A}{8})^2+\frac{A}{8}+\frac{3}{4}] u^{4N+2}\times{\rm id}+\cdots, \label{as-2} \\[4pt]
&&\tilde{t}(u)|_{u\rightarrow \pm\infty}=
-128[\frac32(\frac{A}{8})^2+\frac{A}{16}+\frac{3}{8}]
u^{8N+6}\times {\rm id}+\cdots, \label{as-3}
\eea
where $A$ is a constant given by \bea
A=-2[-2+2(c_1\tilde{c}_1+c_2\tilde{c}_2+c_3\tilde{c}_3)+\frac{c_1c_2\tilde{c}_1\tilde{c}_2}{c_3\tilde{c}_3}
+\frac{c_1c_3\tilde{c}_1\tilde{c}_3}{c_2\tilde{c}_2}+\frac{c_3c_2\tilde{c}_3\tilde{c}_2}{c_1\tilde{c}_1}].
 \label{as-4} \eea
Moreover, we can obtain the values of transfer matrices at some special points:
\bea
&&t(0)=-5\prod_{l=1}^N\rho_{12}(\theta_l)\,\times{\rm  id},\label{sp-1} \\[4pt]
&&t(-1)=-\frac{5}{4}\prod_{l=1}^N(\theta_l-1)(-\theta_l-1)\bar{t}(-\frac12),\label{sp-2} \\[4pt]
&&\tilde{t}(-\frac52)=-\frac{15}{2}\prod_{l=1}^N(\theta_l+1)(-\theta_l+1)(\theta_l+4)(-\theta_l+4)\bar{t}(-\frac72),\label{sp-3} \\[4pt]
&&\tilde{t}(-\frac{13}{2})=330\prod_{l=1}^N(\theta_l-4)(-\theta_l-4){t}(-7),\label{sp-4}\\[4pt]
&&\tilde{t}(-1)=0.\label{sp-5}
\eea
The detail proofs of the above relations are relegated to Appendix D.

\section{Inhomogeneous $T-Q$ relation}
\setcounter{equation}{0}

Since the transfer matrix and the fused ones commutate with each other, they have the common eigenstates.
Acting the crossing relations (\ref{Transfer-Crossing}), (\ref{Crossing-2}) and (\ref{Crossing-3}) and the fusion relations (\ref{Op-pro-open-2}), (\ref{Op-pro-open-3}), (\ref{Op-pro-open-4}), (\ref{Op-pro-open-5}) and (\ref{Op-pro-open-6}) on an eigenstate, we obtain the following functional relations
\bea
&&\hspace{-1.2truecm}\Lambda(u)=\Lambda(-u-6),\quad \bar{\Lambda}(u)=\bar{\Lambda}(-u-6),\quad \tilde{\Lambda}(u)=\tilde{\Lambda}(-u-6),\label{CROSS}\\[6pt]
&&\hspace{-1.2truecm} \Lambda(\theta_j)\Lambda(\theta_j-6)=4^2
\frac{(\theta_j-1)(\theta_j-6)
(\theta_j+1)(\theta_j+6)}{(\theta_j-2)(\theta_j-3)
(\theta_j+2)(\theta_j+3)}\no\\[4pt]
&&\hspace{5mm}\times
(\theta_j-\frac{1}{{2}})(\theta_j-\frac{5}{{2}})
(\theta_j+\frac{1}{{2}})(\theta_j+\frac{5}{{2}})\prod_{i=1}^N{\rho}_{12}(\theta_j-\theta_i){\rho}_{12}(\theta_j+\theta_i),
\label{TQ-1} \\[6pt]
&& \hspace{-1.2truecm}\Lambda(\pm\theta_j)\Lambda(\pm\theta_j-1)=-
\frac{(\pm\theta_j-1)(\pm\theta_j+6)(\pm\theta_j+\frac{5}{{2}})^2
}{(\pm\theta_j+2)(\pm\theta_j+3)
}\no\\[4pt]
&&\hspace{5mm}\times
\prod_{i=1}^N(\pm\theta_j-\theta_i-1)(\pm\theta_j+\theta_i-1)a(\pm\theta_j-\theta_i)a(\pm\theta_j+\theta_i)
\bar{\Lambda}(\pm\theta_j-\frac{1}{{2}}), \label{TQ-2} \\[6pt]
&&\hspace{-1.2truecm}
\Lambda(\pm\theta_j)\bar{\Lambda}(\pm\theta_j-\frac{7}{{2}})=-
\frac{(\pm\theta_j+1)(\pm\theta_j+6) }{(\pm\theta_j-\frac{1}{{2}})
(\pm\theta_j-\frac{3}{{2}})(\pm\theta_j+3)(\pm\theta_j+4)}\no\\[4pt]
&&\hspace{5mm}\times
\prod_{i=1}^N(\pm\theta_j-\theta_i+6)(\pm\theta_j+\theta_i+6)
\tilde{\Lambda}(\pm\theta_j-\frac52), \label{TQ-3} \\[6pt]
&&\hspace{-1.2truecm}
\Lambda(\pm\theta_j)\tilde{\Lambda}(\pm\theta_j\hspace{-0.12truecm}-\hspace{-0.12truecm}\frac92)\hspace{-0.12truecm}=\hspace{-0.12truecm}2^4
\frac{(\pm\theta_j\hspace{-0.12truecm}+\hspace{-0.12truecm}1)(\pm\theta_j\hspace{-0.12truecm}+\hspace{-0.12truecm}6) }{(\pm\theta_j\hspace{-0.12truecm}+\hspace{-0.12truecm}3)
(\pm\theta_j\hspace{-0.12truecm}+\hspace{-0.12truecm}4)}(\pm\theta_j\hspace{-0.12truecm}-\hspace{-0.12truecm}\frac12)
(\pm\theta_j\hspace{-0.12truecm}-\hspace{-0.12truecm}\frac{7}{2})
(\pm\theta_j\hspace{-0.12truecm}+\hspace{-0.12truecm}\frac12)
(\pm\theta_j\hspace{-0.12truecm}+\hspace{-0.12truecm}\frac{5}{2})\no\\[4pt]
&&\hspace{5mm}\times
\prod_{i=1}^N(\pm\theta_j-\theta_i+6)(\pm\theta_j+\theta_i+6)
\Lambda(\pm\theta_j-2)\Lambda(\pm\theta_j-5), \label{TQ-4} \\[6pt]
&&\hspace{-1.2truecm}
\Lambda(\pm\theta_j)\tilde{\Lambda}(\pm\theta_j-\frac{13}{2})=-2^6
\frac{(\pm\theta_j-4)(\pm\theta_j+1)(\pm\theta_j+6)
}{(\pm\theta_j-2)(\pm\theta_j+2)
(\pm\theta_j+3)}\no\\[4pt]
&&\hspace{5mm}\times(\pm\theta_j-\frac{11}{2})(\pm\theta_j-\frac52)(\pm\theta_j-\frac{3}{2})(\pm\theta_j-\frac{1}{2})(\pm\theta_j+\frac12)
(\pm\theta_j+\frac{5}{2})\no\\[4pt]
&&\hspace{5mm}\times
\prod_{i=1}^N(\pm\theta_j-\theta_i+4)(\pm\theta_j+\theta_i+4)a(\pm\theta_j-\theta_i)a(\pm\theta_j+\theta_i)
\Lambda(\pm\theta_j-7), \label{TQ-5}\eea
where $\Lambda(u)$, $\bar{\Lambda}(u)$ and $\tilde{\Lambda}(u)$ are the eigenvalues of transfer matrices
$t(u)$, $\bar{t}(u)$ and $\tilde{t}(u)$, respectively.
Acting the asymptotic behaviors (\ref{as-1})-(\ref{as-3}) on the eigenstate, we have
\bea && \Lambda(u)|_{u\rightarrow \pm\infty}= A   u^{6N+2}+\cdots, \label{as-11} \\[4pt]
&& \bar{\Lambda}(u)|_{u\rightarrow \pm\infty}=  16[-(\frac{A}{8})^2+\frac{A}{8}+\frac{3}{4}] u^{4N+2}+\cdots, \label{as-21} \\[4pt]
&&\tilde{\Lambda}(u)|_{u\rightarrow \pm\infty}=
-128[\frac32(\frac{A}{8})^2+\frac{A}{16}+\frac{3}{8}]
u^{8N+6}+\cdots. \label{as-31} \eea
The constraints (\ref{sp-1})-(\ref{sp-5}) give rise to the relations
\bea
&&\Lambda(0)=-5\prod_{l=1}^N\rho_{12}(\theta_l), \quad  \Lambda(-1)=-\frac{5}{4}\prod_{l=1}^N(\theta_l-1)(-\theta_l-1)\bar{\Lambda}(-\frac12),\label{sp-21} \\[4pt]
&&\tilde{\Lambda}(-\frac52)=-\frac{15}{2}\prod_{l=1}^N(\theta_l+1)(-\theta_l+1)(\theta_l+4)(-\theta_l+4)\bar{\Lambda}(-\frac72),\label{sp-31} \\[4pt]
&&\tilde{\Lambda}(-\frac{13}{2})=330\prod_{l=1}^N(\theta_l-4)(-\theta_l-4){\Lambda}(-7),\quad \tilde{\Lambda}(-1)=0.\label{sp-51} \eea

The $9N+8$ functional relations (\ref{CROSS})-(\ref{sp-51}) allow us completely to determine the eigenvalues, which can be given in terms of some inhomogeneous $T-Q$ relations as
\bea &&\Lambda(u)=\sum_{i=1}^7 Z_i(u)+\sum_{k=1}^2 f_k(u),
\label{Eigen-open-Lambda} \\
&&\bar{\Lambda}(u-\frac12) =-\frac{(u+2)(u+3)}{(u-1)(u+6)(u+\frac52)(u+\frac52)}\prod_{i=1}^N[(u+\theta_i-1)(u-\theta_i-1)a(u-\theta_i)]^{-1} \no\\
&&\qquad \times a^{-1}(u+\theta_i)  \{
Z_1(u)[\sum^{7}_{i=2}Z_i(u-1)+f_1(u-1)+f_2(u-1)]\no\\
&&\qquad +[\sum^{6}_{i=2}Z_i(u)+f_1(u)+f_2(u)]Z_7(u-1)
\no\\[4pt]
&&\qquad+[Z_2(u)+f_1(u)+Z_3(u)][Z_5(u-1)+f_2(u-1)+Z_6(u-1)]\}, \\[4pt]
 &&
\tilde{\Lambda}(u-\frac52) =\frac{u(u-1)(u+4)(u-\frac12)(u-\frac32)}{(u+1)(u+6)(u-4)(u-\frac12)^2}\prod_{i=1}^N[(u+\theta_i-4)(u-\theta_i-4)]^{-1}\no\\[4pt]
&&\qquad\times [(u+\theta_i+6)(u-\theta_i+6)a(u+\theta_i-3)a(u-\theta_i-3)]^{-1} \nonumber\\[4pt]
&&\qquad
\times\{[\sum_{i=1}^4Z_i(u)+f_1(u)][\sum_{k=1}^6Z_k(u-3)+f_1(u-3)+f_2(u-3)]Z_7(u-4)\no\\
&&\qquad+Z_1(u)Z_1(u-3)[\sum_{k=4}^6Z_k(u-4)+f_2(u-4)]+Z_5(u)Z_6(u-3)Z_7(u-4)\no\\[4pt]
&&\qquad+Z_1(u)[Z_2(u-3)+f_1(u-3)+Z_3(u-3)][Z_5(u-4)+f_2(u-4)+Z_6(u-4)]\no\\[4pt]
&&\qquad+Z_1(u)Z_3(u-3)[Z_5(u-4)+f_2(u-4)+Z_6(u-4)]\},\label{Eigen-open-Lambda1} \eea
where the $Z$-functions and $f$-functions are defined as
\bea &&Z_1(u)=- 4\frac{
(u+1)(u+6)}{(u+2)(u+3)
}(u+\frac{1}{{2}})(u+\frac{5}{{2}})\prod_{j=1}^N
a(u-\theta_j)a(u+\theta_j)\frac{Q^{(1)}(u-1)}{Q^{(1)}(u)}\no\\[4pt]
&&Z_2(u)=-4\frac{ u(u+6)}{(u+2)(u+3)
}(u+\frac{1}{{2}})(u+\frac{5}{{2}})\prod_{j=1}^N
b(u-\theta_j)b(u+\theta_j)\frac{Q^{(1)}(u+1)Q^{(2)}(u-3)}{Q^{(1)}(u)Q^{(2)}(u)}\no\\[4pt]
&&Z_3(u)=-4\frac{ u(u+6)}{(u+2)(u+3)
}(u+\frac{7}{{2}})(u+\frac{5}{{2}})\prod_{j=1}^N
b(u-\theta_j)b(u+\theta_j)\frac{Q^{(1)}(u+1)Q^{(2)}(u+3)}{Q^{(1)}(u+3)Q^{(2)}(u)}\no\\[4pt]
&&Z_4(u)=-4\frac{ u(u+6)}{(u+2)(u+4)
}(u+\frac{7}{{2}})(u+\frac{5}{{2}})\prod_{j=1}^N
c(u-\theta_j)c(u+\theta_j)\frac{Q^{(1)}(u+1)Q^{(1)}(u+4)}{Q^{(1)}(u+2)Q^{(1)}(u+3)}\no\\[4pt]
&&Z_5(u)=-4\frac{ u(u+6)}{(u+3)(u+4)
}(u+\frac{7}{{2}})(u+\frac{5}{{2}})\prod_{j=1}^N
d(u-\theta_j)d(u+\theta_j)\frac{Q^{(1)}(u+4)Q^{(2)}(u-1)}{Q^{(1)}(u+2)Q^{(2)}(u+2)}\no\\[4pt]
&&Z_6(u)=-4\frac{ u(u+6)}{(u+3)(u+4)
}(u+\frac{7}{{2}})(u+\frac{11}{{2}})\prod_{j=1}^N
d(u-\theta_j)d(u+\theta_j)\frac{Q^{(1)}(u+4)Q^{(2)}(u+5)}{Q^{(1)}(u+5)Q^{(2)}(u+2)}\no\\[4pt]
&&Z_7(u)=-4\frac{ u(u+5)}{(u+3)(u+4)
}(u+\frac{7}{{2}})(u+\frac{11}{{2}})\prod_{j=1}^N
e(u-\theta_j)e(u+\theta_j)\frac{Q^{(1)}(u+6)}{Q^{(1)}(u+5)},\no \\[4pt]
&&f_1(u)= -4 \frac{u(u+6)}{(u+3) }(u+\frac{5}{{2}})\prod_{j=1}^N
b(u-\theta_j)b(u+\theta_j)\frac{Q^{(1)}(u+1)}{Q^{(2)}(u)}
\,x,\no\\[4pt]
&&f_2(u)=-4 \frac{ u(u+6)}{(u+3) }(u+\frac{7}{2})\prod_{j=1}^N
d(u-\theta_j)d(u+\theta_j)\frac{Q^{(1)}(u+4)}{Q^{(2)}(u+2)} x,
\label{function} \eea $Q$-functions are the standard ones \bea
&&Q^{(1)}(u)=\prod_{k=1}^{L_1}(i u+\mu_k^{(1)}+\frac{i}{2})(i
u-\mu_k^{(1)}+\frac{i}{2}),\no\\
&& Q^{(2)}(u)=\prod_{k=1}^{L_2}(i u+\mu_k^{(2)}+2i)(i
u-\mu_k^{(2)}+2i),\eea
and $x$ is a parameter which will be determined later (see
(\ref{x-value}) below). All the
eigenvalues are the polynomials of $u$, thus the residues of right hand
sides of Eqs.(\ref{Eigen-open-Lambda})-(\ref{Eigen-open-Lambda1}) should be zero, which gives rise to the Bethe ansatz equations (BAEs)
\bea
&&\frac{Q^{(1)}(i\mu_k^{(1)}+\frac{1}{2})Q^{(2)}(i\mu_k^{(1)}-\frac{7}{2})}{Q^{(1)}(i\mu_k^{(1)}-\frac{3}{2})Q^{(2)}(i\mu_k^{(1)}-\frac{1}{2})}
=-\frac{(i\mu_k^{(1)}+\frac{1}{2})}{(i\mu_k^{(1)}-\frac{1}{2})}
\frac{\prod_{j=1}^N(i\mu_k^{(1)}-\theta_j+\frac{1}{2})
(i\mu_k^{(1)}+\theta_j+\frac{1}{2})}{\prod_{j=1}^N(i\mu_k^{(1)}-\theta_j-\frac{1}{2})(i\mu_k^{(1)}+\theta_j-\frac{1}{2})},\no\\
&&\quad k=1,2,\cdots,L_1, \label{opba-1} \\[8pt]
&&\frac{(i\mu_l^{(2)}-\frac{3}{2})}{i\mu_l^{(2)}}
\frac{Q^{(2)}(i\mu_l^{(2)}-5)}{Q^{(1)}(i\mu_l^{(2)}-2)}
+\frac{(i\mu_l^{(2)}+\frac{3}{2})}{i\mu_l^{(2)}}
\frac{Q^{(2)}(i\mu_l^{(2)}+1)}{Q^{(1)}(i\mu_l^{(2)}+1)}\no\\[4pt]
&&\qquad\qquad=-x, \quad l=1,2,\cdots,L_2. \label{opba-2}\eea From
the asymptotic behaviors of $\Lambda(u)$, $\bar{\Lambda}(u)$ and
$\tilde{\Lambda}(u)$, we obtain the constraint between the
integers $L_1$ and $L_2$ i.e., $L_2=L_1$. Moreover, the value of
parameter $x$ in the functions $f_1(u)$ and $f_2(u)$ is \bea
&&x=\frac14\Big[-16+2(c_1\tilde{c}_1+c_2\tilde{c}_2+c_3\tilde{c}_3)+\frac{c_1c_2\tilde{c}_1\tilde{c}_2}{c_3\tilde{c}_3}
+\frac{c_1c_3\tilde{c}_1\tilde{c}_3}{c_2\tilde{c}_2}+\frac{c_3c_2\tilde{c}_3\tilde{c}_2}{c_1\tilde{c}_1}\Big].\label{x-value}
\eea
 According to
(\ref{K-matrix-dig1}) and (\ref{ksk}), while
$c_1=0,c_2=0,c_3=2,\tilde{c}_1=0,\tilde{c}_2=0,\tilde{c}_3=2$, we
have  $x=0$ from  (\ref{x-value}). Referring to (\ref{q}) and
(\ref{q1}), that means (\ref{Eigen-open-Lambda}) just the result
obtained in \cite{yung} when taking rational limit.

We have done the numerical computation  with some small sites for the above BAEs (\ref{opba-1})-(\ref{x-value}). The results with $N=1$ and $N=2$ are shown in Table \ref{roots_real3} and \ref{roots_real} (see Appendix E), which give the same
complete sets of the eigenvalues $\Lambda(u)$ by (\ref{Eigen-open-Lambda}) as those obtained by directly diagonalizing the transfer matrix.

\section{Results for the periodic boundary condition}
\setcounter{equation}{0}

For the periodic boundary condition case, the  transfer matrix and the fused ones are
\bea
t^{(p)}(u)=tr_0T_0(u), \quad
\bar{t}^{(p)}(u)=tr_{\bar{0}}T_{\bar{0}}(u), \quad
\tilde{t}^{(p)}(u)=tr_{\tilde{0}}T_{\tilde{0}}(u). \eea
Using the similar method, we can show that they satisfy the closed operators product identities
\bea &&
t^{(p)}(\theta_j)\,t^{(p)}(\theta_j-6)=\prod_{i=1}^N
a(\theta_j-\theta_i)e(\theta_j-\theta_i-6)\times {\rm id},\label{Op-Product-Periodic-1}  \\
&& t^{(p)}(\theta_j)\,t^{(p)}(\theta_j-4)= \prod_{i=1}^N
(\theta_j-\theta_i+1)(\theta_j-\theta_i-4)(\theta_j-\theta_i-6)\,t^{(p)}(\theta_j-2),\label{Op-Product-Periodic-2} \\
&& t^{(p)}(\theta_j)\,t^{(p)}(\theta_j-1)= \prod_{i=1}^N
(\theta_j-\theta_i-1)a(\theta_j-\theta_i)\,\bar{t}^{(p)}(\theta_j-\frac12),\label{Op-Product-Periodic-3}  \\
&& t^{(p)}(\theta_j)\,\bar{t}^{(p)}(\theta_j-\frac{11}{2})=
\prod_{i=1}^N
(\theta_j-\theta_i+4)(\theta_j-\theta_i+6)\,t^{(p)}(\theta_j-5),\label{Op-Product-Periodic-4}  \\
&& t^{(p)}(\theta_j)\,\bar{t}^{(p)}(\theta_j-\frac{7}{2})=
\prod_{i=1}^N(\theta_j-\theta_i+6)\,\tilde{t}^{(p)}(\theta_j-\frac52),\label{Op-Product-Periodic-5}  \\
&& t^{(p)}(\theta_j)\,\tilde{t}^{(p)}(\theta_j-\frac72)=
\prod_{i=1}^N
(\theta_j-\theta_i-1)(\theta_j-\theta_i-4)a(\theta_j-\theta_i)\,\bar{t}^{(p)}(\theta_j-\frac52),\label{Op-Product-Periodic-6}  \\
&& t^{(p)}(\theta_j)\,\tilde{t}^{(p)}(\theta_j-\frac92)=
\prod_{i=1}^N
(\theta_j-\theta_i+6)\,t^{(p)}(\theta_j-2)t^{(p)}(\theta_j-5),\label{Op-Product-Periodic-7}  \\
&& t^{(p)}(\theta_j)\,\tilde{t}^{(p)}(\theta_j-\frac{13}{2})=
\prod_{i=1}^N
(\theta_j-\theta_i-4)a(\theta_j-\theta_i)\,t^{(p)}(\theta_j-7),\label{Op-Product-Periodic-8}  \\
&& t^{(p)}(\theta_j)\,\tilde{t}^{(p)}(\theta_j-\frac32)=
\prod_{i=1}^N
(\theta_j-\theta_i-1)(\theta_j-\theta_i-6)\,t^{(p)}(\theta_j-2)\bar{t}^{(p)}(\theta_j-\frac12).\label{Op-Product-Periodic-9}
\eea
Moreover, the asymptotic behaviors of transfer matrices become
\bea && t^{ (p)}(u)|_{u\rightarrow \pm\infty}= 7u^{3N}\times {\rm  id} +\cdots,\quad
\bar{t}^{(p)}(u)|_{u\rightarrow \pm\infty}=
15u^{2N}\times {\rm  id} +\cdots, \\[4pt]
&& \tilde{t}^{(p)}(u)|_{u\rightarrow \pm\infty}= 34u^{4N}\times
{\rm id} +\cdots.\label{fuwwwtpl-7} \eea

From the definitions, we know that the transfer matrices $ t^{ (p)}(u)$, $\bar{t}^{(p)}(u)$ and $\tilde{t}^{(p)}(u)$ are the polynomials
of $u$ with degrees $3N$, $2N$ and $4N$, respectively. Thus their eigenvalues can be determined by $9N+3$ independent conditions.
The constraints (\ref{Op-Product-Periodic-1})-(\ref{fuwwwtpl-7}) give us sufficient
information to obtain these eigenvalues. Denote the eigenvalues of $t^{(p)}(u)$, $\bar{t}^{(p)}(u)$ and
$\tilde{t}^{(p)}(u)$ as $\Lambda^{(p)}(u)$, $\bar{\Lambda}^{(p)}(u)$ and
$\tilde{\Lambda}^{(p)}(u)$, respectively. Then we can express the eigenvalues in terms of the homogeneous $T-Q$ relations
\bea
&&\Lambda^{(p)}(u)=\sum_{j=1}^7 Z^{(p)}_j(u), \label{T-Q-Hom-1} \\[4pt]
&&\Lambda^{(p)}_2(u-\frac12)=\prod_{i=1}^N
((u-\theta_i-1)a(u-\theta_i))^{-1}\,\{
Z^{(p)}_1(u)[\sum_{j=2}^7Z^{(p)}_j(u-1)]\no\\
&&\qquad +[\sum_{j=2}^6Z^{(p)}_j(u)]Z^{(p)}_7(u-1)+[Z^{(p)}_2(u)+Z^{(p)}_3(u)][Z^{(p)}_5(u-1)+Z^{(p)}_6(u-1)]\},\label{T-Q-Hom-2} \\[4pt]
&&\Lambda^{(p)}_3(u-\frac52)=\prod_{i=1}^N
((u-\theta_i-4)(u-\theta_i+6)a(u-\theta_i-3))^{-1}\no\\[4pt]
&&\qquad \times
\{(\sum_{j=1}^4Z^{(p)}_j(u))(\sum_{k=1}^6Z^{(p)}_j(u-3))Z^{(p)}_7(u-4)+Z^{(p)}_1(u)Z^{(p)}_1(u-3)[\sum_{j=4}^6Z^{(p)}_j(u-4)]\no\\[4pt]
&&\qquad
+Z^{(p)}_1(u)[Z^{(p)}_2(u-3)+Z^{(p)}_3(u-3)][Z^{(p)}_5(u-4)+Z^{(p)}_6(u-4)]+Z^{(p)}_2(u)Z^{(p)}_3(u-3)\no\\[4pt]
&&\qquad \times [Z^{(p)}_5(u-4)+Z^{(p)}_6(u-4)]+Z^{(p)}_5(u)Z^{(p)}_6(u-3)Z^{(p)}_7(u-4)\},\label{T-Q-Hom-3} \eea
where the $Z$-functions are \bea
&&Z^{(p)}_1(u)=\prod_{j=1}^N
a(u-\theta_j)\,\frac{Q_{p}^{(1)}(u-1)}{Q_{p}^{(1)}(u)},\no\\[4pt]
&&Z^{(p)}_2(u)=\prod_{j=1}^Nb(u-\theta_j)\,\frac{Q_{p}^{(1)}(u+1)Q_{p}^{(2)}(u-3)}{Q_{p}^{(1)}(u)Q_{p}^{(2)}(u)},\no\\[4pt]
&&Z^{(p)}_3(u)=\prod_{j=1}^N
b(u-\theta_j)\,\frac{Q_{p}^{(1)}(u+1)Q_{p}^{(2)}(u+3)}{Q_{p}^{(1)}(u+3)Q_{p}^{(2)}(u)},\no\\[4pt]
&&Z^{(p)}_4(u)=\prod_{j=1}^N
c(u-\theta_j)\,\frac{Q_{p}^{(1)}(u+1)Q_{p}^{(1)}(u+4)}{Q_{p}^{(1)}(u+2)Q_{p}^{(1)}(u+3)},\no\\
&&Z^{(p)}_5(u)=\prod_{j=1}^N d(u-\theta_j)\,
\frac{Q_{p}^{(1)}(u+4)Q_{p}^{(2)}(u-1)}{Q_{p}^{(1)}(u+2)Q_{p}^{(2)}(u+2)},\no\\[4pt]
&&Z^{(p)}_6(u)=\prod_{j=1}^N d(u-\theta_j)\,
\frac{Q_{p}^{(1)}(u+4)Q_{p}^{(2)}(u+5)}{Q_{p}^{(1)}(u+5)Q_{p}^{(2)}(u+2)},\no\\[4pt]
&&Z^{(p)}_7(u)=\prod_{j=1}^N e(u-\theta_j)\,
\frac{Q_{p}^{(1)}(u+6)}{Q_{p}^{(1)}(u+5)},\no\\[4pt]
&&Q_{p}^{(1)}(u)=\prod_{k=1}^{L_1}(i
u+\mu_k^{(1)}+i\frac{1}{2}),\quad
Q_{p}^{(2)}(u)=\prod_{k=1}^{L_2}(i u+\mu_k^{(2)}+2i).\eea
The regularity of the expressions (\ref{T-Q-Hom-1})-(\ref{T-Q-Hom-3}) of  eigenvalues
requires that the
Bethe roots $\{\mu^{(m)}_k\}$ should satisfy the BAEs
\bea &&
\frac{Q_{p}^{(1)}(i\mu_k^{(1)}+\frac{1}{2})Q_{p}^{(2)}(i\mu_k^{(1)}-\frac{7}{2})}{Q_{p}^{(1)}(i\mu_k^{(1)}-\frac{3}{2})Q_{p}^{(2)}(i\mu_k^{(1)}-\frac{1}{2})}
=-\prod_{j=1}^N \frac{i\mu_k^{(1)}+\frac{1}{2}-\theta_j
}{i\mu_k^{(1)}-\frac{1}{2}-\theta_j}, \quad k=1,\cdots, L_1, \label{BAEs-1} \\[8pt]
&&\frac{Q_{p}^{(1)}(i\mu_l^{(2)}-2)Q_{p}^{(2)}(i\mu_l^{(2)}+1)}{Q_{p}^{(1)}(i\mu_l^{(2)}+1)Q_{p}^{(2)}(i\mu_l^{(2)}-5)}
=-1, \quad l=1,\cdots, L_2.\label{BAEs-2} \eea
We have verified that the above BAEs indeed guarantee all the $T-Q$ relations
(\ref{T-Q-Hom-1})-(\ref{T-Q-Hom-3}) are the polynomials of $u$ with
the required degrees. Moreover, we have checked that our result (\ref{T-Q-Hom-1}) and the associated BAEs (\ref{BAEs-1})-(\ref{BAEs-2}) coincide with those obtained in
\cite{martins}.

\section{Conclusions}

In conclusion,  the exact solution of the $G_2$ quantum integrable spin chain is studied by introducing a new non-diagonal boundary condition. A closed set of fusion identities are derived, which allow us to determine the energy spectrum and Bethe ansatz
equations in an analytic way.
It  demonstrates that the off-diagonal Bethe ansatz method \cite{Cao1,Cao13,Cao14,Hao14,Cao2}  is also applicable  to
integrable models associated with the exceptional Lie algebras.

\section*{Acknowledgments}

The financial supports from National Key R$\&$D Program of China (Grant No.2021YFA1402104),
the National Natural Science Foundation of China (Grant Nos. 12434006, 12247103, 12205235, 12147160, 12074410, 12075177 and 11934015), Major Basic Research Program of Natural Science of Shaanxi Province (Grant Nos. 2021JCW-19 and 2017ZDJC-32), Shaanxi Fundamental Science Research Project for Mathematics
and Physics (Grant Nos. 23JS0008 and 22JSZ005) and Strategic Priority Research Program of the Chinese Academy of Sciences (Grant No. XDB33000000) are gratefully acknowledged.


\section*{Appendix A: Fusions of the R-matrices and K-matrices}
\setcounter{equation}{0}
\renewcommand{\theequation}{A.\arabic{equation}}
\subsection*{Appendix A.1: Fusions of the fundamental $R$-matrix and $K$-matrix }
The $R$-matrix (\ref{rm}) can also be written in terms of the projectors as
 \bea R_{12}(u)&=&(u-1)(u+4)(u-6)P_{12}^{(1)}+(u+1)(u-4)(u+6)P_{12}^{(7)}\no\\[4pt]
 &&+(u-1)(u+4)(u+6)P_{12}^{(14)}+(u+1)(u+4)(u+6)P_{12}^{(27)}
,\label{C-1S1}\eea where $P_{12}^{(d)}$ are $d$-dimensional
projectors, where $d=1, 7, 14, 27$.  Thus the $R$-matrix can
degenerate into the projectors at
certain points of the spectral
parameter. For an example, if $u=-6$, we have
\bea R_{12}(-6)= P^{(1)
}_{12}\times S_1, \label{pro-0}
\eea where $S_1$ is an irrelevant
constant matrix omitted here,  $P^{(1)}_{12}$ is the
$1$-dimensional projector \bea
P^{(1)}_{12}=|\psi_0\rangle\langle\psi_0|, \quad  P^{(1) }_{21}=
P^{(1) }_{12}, \label{a1} \eea the vector $|\psi_0\rangle$ is
\bea
|\psi_0\rangle=\frac{1}{\sqrt{7}}(|17\rangle-|26\rangle+|35\rangle-|44\rangle+|53\rangle-|62\rangle+|71\rangle).\label{1-dim}
\eea
When $u=-1$, we have
\bea R_{12}(-1)= P^{(15) }_{12}\times S_{15}, \label{pro-15}\eea
where $S_{15}$ is an irrelevant constant matrix omitted here,
$P^{(15)}_{12}$ is the 15-dimensional projector \bea P^{(15)
}_{12}=\sum_{i=1}^{15}|\psi_i^{(15)}\rangle\langle\psi_i^{(15)}|,\label{a15}
\eea and the related vectors are \bea
&&|\psi_1^{(15)}\rangle=\frac{1}{\sqrt{2}}(|12\rangle-|21\rangle),\ |\psi_2^{(15)}\rangle=\frac{1}{\sqrt{2}}(|13\rangle-|31\rangle),\no\\[4pt]
&&|\psi_3^{(15)}\rangle=\frac{1}{\sqrt{3}}(|14\rangle-|41\rangle)+\frac{1}{\sqrt{6}}(|23\rangle-|32\rangle),\no\\[4pt]
&&|\psi_4^{(15)}\rangle=\frac{1}{\sqrt{3}}(|15\rangle-|51\rangle)+\frac{1}{\sqrt{6}}(|24\rangle-|42\rangle),\no\\[4pt]
&&|\psi_5^{(15)}\rangle=\frac{1}{\sqrt{3}}(|16\rangle-|61\rangle)+\frac{1}{\sqrt{6}}(|34\rangle-|43\rangle),\no\\[4pt]
&&|\psi_6^{(15)}\rangle=\frac{1}{\sqrt{19}}(|17\rangle-3|71\rangle+|44\rangle-2|53\rangle+2|62\rangle),\ |\psi_7^{(15)}\rangle=\frac{1}{\sqrt{2}}(|25\rangle-|52\rangle),\no\\[4pt]
&&|\psi_8^{(15)}\rangle=\frac{1}{\sqrt{190}}(-\frac{13}{2}|17\rangle+\frac{19}{2}|26\rangle+3|44\rangle-6|53\rangle-\frac{7}{2}|62\rangle+\frac{1}{2}|71\rangle),\no\\[4pt]
&&|\psi_9^{(15)}\rangle=\frac{1}{\sqrt{3}}(|45\rangle-|54\rangle)+\frac{1}{\sqrt{6}}(|27\rangle-|72\rangle),\no\\[4pt]
&&|\psi_{10}^{(15)}\rangle=\frac{1}{\sqrt{210}}(-\frac{13}{2}|17\rangle+\frac{1}{2}|26\rangle+10|35\rangle
-3|44\rangle-4|53\rangle-\frac{13}{2}|62\rangle-\frac{1}{2}|71\rangle),\no\\[4pt]
&&|\psi_{11}^{(15)}\rangle=\frac{1}{\sqrt{2}}(|36\rangle-|63\rangle),\ |\psi_{12}^{(15)}\rangle=\frac{1}{\sqrt{3}}(|46\rangle-|64\rangle)+\frac{1}{\sqrt{6}}(|37\rangle-|73\rangle),\no\\[4pt]
&&|\psi_{13}^{(15)}\rangle=\frac{1}{\sqrt{3}}(|47\rangle-|74\rangle)+\frac{1}{\sqrt{6}}(|56\rangle-|65\rangle),\no\\[4pt]
&&|\psi_{14}^{(15)}\rangle=\frac{1}{\sqrt{2}}(|57\rangle-|75\rangle),\
|\psi_{15}^{(15)}\rangle=\frac{1}{\sqrt{2}}(|67\rangle-|76\rangle).
\eea The projectors $P^{(1)}_{21}$ and $P^{(15)}_{21}$ can be obtained by exchanging
two spaces $V_{1}$ and $V_{2}$., i.e., $|kl\rangle\rightarrow
|lk\rangle$.

The fusion with
15-dimensional projector $P_{12}^{(15)}$ from (\ref{pro-15}) gives \bea
&&P^{(15)}_{12}R_{23}(u)R_{13}(u-1)P^{(15)}_{12}=(u-1)(u+1)(u+4)(u+6)R_{\bar{1}3}(u-\frac12),\no\\[4pt]
&&P^{(15)}_{21}R_{32}(u)R_{31}(u-1)P^{(15)}_{21}=(u-1)(u+1)(u+4)(u+6)R_{3\bar{1}}(u-\frac12),\label{1YBE15}
\eea
where the subscript $\bar{1}$ means the 15-dimensional fused space and $R_{\bar{1}3}(u)$ is the $(15\times 7)\times (15\times 7)$-dimensional
fused $R$-matrix. The matrix elements of
$R_{\bar{1}3}(u)$ are the polynomials of $u$, and the maximum degree is 2.
The fused $R$-matrix (\ref{1YBE15}) has the properties
\bea
&&R_{\bar{1}2}(u)R_{2\bar{1}}(-u)={\rho}_{\bar{1}2}(u)=(u+\frac{7}{2})(u+\frac{11}{2})(u-\frac{7}{2})(u-\frac{11}{2}),\label{C-S-1}\\[4pt]
&&R_{\bar{1}2}(u)^{t_{\bar{1}}}R_{2\bar{1}}(-u-12)^{t_{\bar{1}}}=\tilde{\rho}_{\bar{1}2}(u)
=\rho_{\bar{1}2}(u+6),\no\\[4pt]
&&R_{\bar{1}2}(u)=V_{\bar{1}}R^{t_{\bar{1}}}_{2\bar{1}}(-u-6)[V_{\bar{1}}]^{-1}, \quad V_{\bar{1}}=P^{(15)}_{12}V_{2}V_{1}P^{(15)}_{12}, \no\\[4pt]
&&R_{2\bar{1}}(u)=V_{\bar{1}}^{t_{\bar{1}}}R^{t_{\bar{1}}}_{\bar{1}2}(-u-6)[V_{\bar{1}}^{t_{\bar{1}}}]^{-1}.
\label{cr-1}
\eea
and satisfies the quantum Yang-Baxter equation
  \bea  R_{{\bar 1}2}(u-v)  R_{{\bar 1}3}(u) R_{{2}3}(v)=R_{{2}3}(v)
R_{{\bar 1}3}(u)   R_{{\bar 1}2}(u-v). \label{YBE15}
\eea
The 15-dimensional fusion of reflection matrices gives
\bea
&&\hspace{-2.2truecm}P^{(15)}_{12}K^-_2(u)R_{12}(2u-1)K^-_1(u-1)P^{(15)}_{21}\no\\[4pt]
&&\hspace{-0.2truecm}=8(u-\frac12)(u+\frac12)(u+\frac52)(u-1)K^-_{\bar{1}}(u-\frac{1}{2}),\no\\[6pt]
&&\hspace{-2.2truecm}P^{(15)}_{21}K^+_1(u-1)R_{21}(-2u-2\kappa+1)K^+_2(u)P^{(15)}_{12}\no\\[4pt]
&&\hspace{-0.2truecm}=8(u+\frac{5}{2})(u+\frac{9}{2})(u+\frac{11}{2})(u+6)K^+_{\bar{1}}(u-\frac{1}{2}), \label{2YBE15}
\eea
where $K^{\mp}_{\bar{1}}(u)$ are the $15\times 15$-dimensional fused reflection matrices.
The matrix elements of
$K^{\mp}_{\bar{1}}(u)$ are the polynomials of $u$ and the maximum degree is 1. Moreover, $K^{\mp}_{\bar{1}}(u)$ satisfy the
reflection equations
\begin{eqnarray}
&&  R _{\bar 12}(u-v) K^{-}_{\bar  1}(u)  R _{2\bar 1}(u+v)
K^{-}_{2}(v)=K^{-}_{2}(v)
 R_{\bar 12}(u+v)  K^{-}_{\bar 1}(u)  R _{2\bar 1}(u-v), \label{RE15-1}\\[6pt]
&&  R _{\bar 12}(-u+v)  K^{+}_{\bar 1}(u)  R _{2\bar 1}
 (-u-v-12) K^{ +}_{2}(v) \nonumber\\[4pt]
&&\qquad\qquad\qquad\qquad =K^{ +}_{2}(v)  R _{\bar 12}(-u-v-12)  K^{ +}_{\bar 1}(u)
R _{2\bar 1}(-u+v). \label{RE15-2}
\end{eqnarray}

\subsection*{Appendix A.2: Fusions of the fused $R$-matrix and $K$-matrix }

The fused $R$-matrix (\ref{1YBE15}) also has the degenerated points. For example,
\bea && R_{\bar{1}2}(-\frac72)= P^{(34) }_{\bar{1}2}\times
S_{34},\label{C-34} \eea
where $S_{34}$ is an irrelevant constant matrix and $P_{\bar{1}2}^{(34)}$ is a
34-dimensional projector, which allows us to take the fusion again.
Repeating the similar processes, we obtain the next fused $R$-matrices as
\bea
&&P^{(34)}_{\bar{1}2}R_{23}(u)R_{\bar{1}3}(u-\frac72)P^{(34)}_{\bar{1}2}=(u+6)R_{\tilde{1}3}(u-\frac52), \no \\[6pt]
&&P^{(34)}_{2\bar{1}}R_{32}(u)R_{3\bar{1}}(u-\frac72)P^{(34)}_{2\bar{1}}
=(u+6)Q_{\tilde{1}}R_{3\tilde{1}}(u-\frac52)Q_{\tilde{1}}^{-1},\label{oip1}
\eea where the subscript $\tilde{1}$ denotes the 34-dimensional fused space ${\bf V}_{\langle 2\bar 1\rangle} $ and $Q_{\tilde{1}}$ is a $34\times 34$ matrix defined in the fused space. We note that
$P_{2\bar{1}}^{(34)}\ne P_{\bar{1}2}^{(34)}$. The next fused $R$ matrix (\ref{oip1}) is a $34^2 \times 7^2$ matrix thus the detailed form is omitted here. The matrix elements of $R_{\tilde{1}2}(u)$
are the polynomials of $u$ and the maximum degree of these polynomials is 4. The next fused $R_{\tilde{1}2}(u)$
matrix has the following properties
\begin{eqnarray}
 &&R_{\tilde{1}2}(u)R_{2\tilde{1}}(-u)= (u^2-\frac{9}{4})(u^2-\frac{49}{4})(u^2-\frac{81}{4})(u^2-\frac{169}{4})\equiv{\rho}_{\tilde{1}2}(u),\no\\
&&R_{\tilde{1}2}(u)^{t_{\tilde{1}}}R_{2\tilde{1}}(-u-12)^{t_{\tilde{1}}}=\rho_{\tilde{1}2}(u+6)\equiv\tilde{\rho}_{\tilde{1}2}(u),\label{C-S-4}
\end{eqnarray}
and satisfies the Yang-Baxter equation \bea R_{{\tilde 1}2}(u-v)  R_{{\tilde 1}3}(u)
R_{{2}3}(v)=R_{{2}3}(v) R_{{\tilde 1}3}(u)   R_{{\tilde
1}2}(u-v)\label{YBE34}. \eea

The related next fused reflection matrices are obtained by taking the fusion of reflection matrices
with the 34-dimensional projectors as
\bea
&&\hspace{-1.2truecm}P^{(34)}_{\bar{1}2}K^-_2(u)R_{\bar{1}2}(2u-\frac72)K^-_{\bar{1}}(u-\frac72)P^{(34)}_{2\bar{1}}=4(u+1)K^-_{\tilde{1}}(u-\frac52)Q_{\tilde{1}}^{-1},\\[6pt]
&&\hspace{-1.2truecm}P^{(34)}_{2\bar{1}}K^+_{\bar{1}}(u-\frac72)R_{2\bar{1}}(-2u-2\kappa+\frac72)K^+_2(u)P^{(34)}_{\bar{1}2}
=-4(u+6)Q_{\tilde{1}}K^+_{\tilde{1}}(u-\frac52), \eea
where all the matrix elements of $K^{\mp}_{\tilde{1}}(u)$ are the polynomials of $u$, and among of them the maximum degree of these polynomials is 3.
The next fused reflection matrices satisfy the reflection equation
\begin{eqnarray}
&&  R _{\tilde 12}(u-v) K^{-}_{\tilde  1}(u) R _{2\tilde 1}(u+v)
K^{-}_{2}(v)=K^{-}_{2}(v)
 R_{\tilde 12}(u+v)  K^{-}_{\tilde 1}(u)  R _{2\tilde 1}(u-v), \label{RE34-1}\\[6pt]
&&  R _{\tilde 12}(-u+v)  K^{+}_{\tilde 1}(u)  R _{2\tilde 1}
 (-u-v-12) K^{ +}_{2}(v) \nonumber\\[4pt]
&&\qquad\qquad =K^{ +}_{2}(v) R _{\tilde 12}(-u-v-12)  K^{ +}_{\tilde
1}(u) R _{2\tilde 1}(-u+v). \label{RE34-2}
\end{eqnarray}


\section*{Appendix B: Proofs of the crossing relations}
\setcounter{equation}{0}
\renewcommand{\theequation}{B.\arabic{equation}}
\subsection*{Appendix B.1: Proof of (\ref{Transfer-Crossing})}
Let us show that the transfer matrix (\ref{t-1}) possesses the crossing symmetry (\ref{Transfer-Crossing}).
With the help of crossing symmetry (\ref{C-S}) of $R$-matrix, the transposition in the auxiliary space of single-row monodromy
matrix $T_0(u)$ satisfies \bea
&&T_0^{t_0}(-u-6)=\{R_{01}(-u-6-\theta_1)R_{02}(-u-6-\theta_{2})\cdots
R_{0N}(-u-6-\theta_N)\}^{t_0}\no\\[6pt]
&&\hspace{10mm}=(-1)^N\{V_0R_{10}^{t_0}(u+\theta_1)R_{20}^{t_{0}}(u+\theta_{2})\cdots
R_{N0}^{t_0}(u+\theta_N)V_0^{-1}\}^{t_0}\no\\[6pt]
&&\hspace{10mm}=(-1)^N[V_0^{t_0}]^{-1}\{R_{N0}(u+\theta_N)R_{N-10}(u+\theta_{N-1})\cdots
R_{10}(u+\theta_1)\}V_0^{t_0}\no\\[6pt]
 &&\hspace{10mm} =(-1)^N[V_0^{t_0}]^{-1}\hat{T}_0(u)V_0^{t_0}.\label{kcm}\eea
Similarly, we have $\hat{T}_0^{t_0}(-u-6)=(-1)^N
V_0^{-1}{T}_0(u)V_0$. The direct calculation gives \bea
&&tr_1 \{{R}_{12}(0)R_{12}(2u)V_1[K_1^-(-u-6)]^{t_1}[V_1^{t_1}]^{-1}\}=f(u)K^-_2(u),\no \\[4pt]
&&
tr_2\{{R}_{12}(0)R_{12}(2u)K_2^+(u)\}=f(u)V_1^{t_1}K^+_1(-u-6)^{t_1}V_1^{-1},\label{kcp1}
\eea where $f(u)=-96(u+1)(u+6)(2u+5)$. Combining the results of
Eqs.(\ref{kcm})-(\ref{kcp1}), we obtain \bea
&&t(-u-6)=tr_0\{K^+_0(-u-6)T_0(-u-6)\}^{t_0}\{K^-_0(-u-6)\hat{T}_0(-u-6)\}^{t_0}\no\\
&&\hspace{10mm}=tr_0\hat{T}_0(u)V_0^{t_0}\{K^+_0(-u-6)\}^{t_0}V_0^{-1}T_0(u)V_0\{K^-_0(-u-6)\}^{t_0}[V_0^{t_0}]^{-1}\no\\[4pt]
&&\hspace{10mm}=tr_0\hat{T}_0(u)tr_1{R}_{01}(0)R_{01}(2u)K^+_1(u)T_0(u)V_0\{K^-_0(-u-6)\}^{t_0}[V_0^{t_0}]^{-1}/f(u)\no\\[4pt]
&&\hspace{10mm}=tr_1tr_0{R}_{10}(0)\hat{T}_1(u)R_{01}(2u)T_0(u)V_0\{K^-_0(-u-6)\}^{t_0}[V_0^{t_0}]^{-1}K^+_1(u)/f(u)\no\\[4pt]
&&\hspace{10mm}=tr_1tr_0{R}_{10}(0)T_0(u)R_{01}(2u)\hat{T}_1(u)V_0\{K^-_0(-u-6)\}^{t_0}[V_0^{t_0}]^{-1}K^+_1(u)/f(u)\no\\[4pt]
&&\hspace{10mm}=tr_1T_1(u)tr_0{R}_{01}(0)R_{01}(2u)V_0\{K^-_0(-u-6)\}^{t_0}[V_0^{t_0}]^{-1}\hat{T}_1(u)K^+_1(u)/f(u)\no\\[4pt]
&&\hspace{10mm}=tr_1K^+_1(u)T_1(u)K^-_1(u)\hat{T}_1(u)=t(u),\label{n-13}\eea
where we have used the following relations
\bea
&&\hat{T}_1(u)R_{01}(2u)T_0(u)=T_0(u)R_{01}(2u)\hat{T}_1(u),\\[6pt]
&&{R}_{10}(0)T_0(u)=T_1(u){R}_{01}(0),\quad
\hat{T}_0(u){R}_{01}(0)={R}_{10}(0)\hat{T}_1(u). \eea

\subsection*{Appendix B.2: Proof of (\ref{Crossing-2})}

We prove that the fused transfer matrix $\bar{t}(u)$ satisfies the crossing symmetry (\ref{Crossing-2})
\bea
\bar{t}(-u-6)=\bar{t}(u).
\eea
For this purpose, we need take the fusion of $R_{0\bar{1}}(u)$ (\ref{1YBE15}) in the quantum space
by the 15-dimensional projector and the result is
\bea
P^{(15)}_{01}R_{1\bar{1}}(u+\frac12)R_{0\bar{1}}(u-\frac12)P^{(15)}_{01}= R_{\bar{0}\bar{1}}(u).\label{kcp21}
\eea
The fused $R$-matrix (\ref{kcp21}) satisfies the Yang-Baxter equation.
\bea
\hat{T}_{\bar{1}}(u)R_{\bar{0}\bar{1}}(u+v)T_{\bar{0}}(v)=T_{\bar{0}}(v)R_{\bar{0}\bar{1}}(u+v)
\hat{T}_{\bar{1}}(u).\label{cc-23}
\eea
At the point of $u=0$, the fused $R$-matrix (\ref{kcp21}) reduces to the permutation
operator, which leads to
\bea
\hat{T}_{\bar{0}}(u){R}_{\bar{0}\bar{1}}(0)={R}_{\bar{1}\bar{0}}(0)\hat{T}_{\bar{1}}(u),\qquad
{R}_{\bar{1}\bar{0}}(0)T_{\bar{0}}(u)=T_{\bar{1}}(u){R}_{\bar{0}\bar{1}}(0).\label{ding}
\eea
With the help of Eq.(\ref{cr-1}), the transposition of monodromy
matrix $T_{\bar{0}}(u)$ in the auxiliary space satisfies \bea
&&T_{\bar{0}}^{t_{\bar{0}}}(-u-6)=\{R_{{\bar{0}}1}(-u-6-\theta_1)R_{{\bar{0}}2}(-u-6-\theta_{2})\cdots
R_{{\bar{0}}N}(-u-6-\theta_N)\}^{t_{\bar{0}}}\no\\[4pt]
&&\hspace{10mm}=\{V_{\bar{0}}R_{1{\bar{0}}}^{t_{\bar{0}}}(u+\theta_1)R_{2{\bar{0}}}^{t_{{\bar{0}}}}(u+\theta_{2})\cdots
R_{N{\bar{0}}}^{t_{\bar{0}}}(u+\theta_N)V_{\bar{0}}^{-1}\}^{t_{\bar{0}}}\no\\[4pt]
&&\hspace{10mm}=[V_{\bar{0}}^{t_{\bar{0}}}]^{-1}\{R_{N{\bar{0}}}(u+\theta_N)R_{N-1{\bar{0}}}(u+\theta_{N-1})\cdots
R_{1{\bar{0}}}(u+\theta_1)\}V_{\bar{0}}^{t_{\bar{0}}}\no\\[4pt]
 &&\hspace{10mm} =[V_{\bar{0}}^{t_{\bar{0}}}]^{-1}\hat{T}_{\bar{0}}(u)V_{\bar{0}}^{t_{\bar{0}}}.\label{kcm-1}\eea
Similarly, we have \bea&&\hat{T}_{\bar{0}}^{t_{\bar{0}}}(-u-6)=
V_{\bar{0}}^{-1}{T}_{\bar{0}}(u)V_{\bar{0}}.\label{kcm-2}\eea
By using the rules of taking trace, we have
\bea
\bar{t}(-u-6)=tr_{\bar{0}}\{K^+_{\bar{0}}(-u-6)
T_{\bar{0}}(-u-6)\}^{t_{\bar{0}}}\{K^-_{\bar{0}}(-u-6)\hat{T}_{\bar{0}}(-u-6)\}^{t_{\bar{0}}}.\label{cc-1}
\eea
Substituting Eqs.(\ref{kcm-1}) and (\ref{kcm-2}) into (\ref{cc-1}), we obtain
\bea
\bar{t}(-u-6)=tr_{\bar{0}}\hat{T}_{\bar{0}}(u)V_{\bar{0}}^{t_{\bar{0}}}
\{K^+_{\bar{0}}(-u-6)\}^{t_{\bar{0}}}V_{\bar{0}}^{-1}
T_{\bar{0}}(u)V_{\bar{0}}\{K^-_{\bar{0}}(-u-6)\}^{t_{\bar{0}}}[V_{\bar{0}}^{t_{\bar{0}}}]^{-1}.
\label{cc-2}
\eea
The fused reflection matrix satisfies
\bea
V_{\bar{0}}^{t_{\bar{0}}}
K^+_{\bar{0}}(-u-6)^{t_{\bar{0}}}V_{\bar{0}}^{-1}
=\bar{f}^{-1}(u)tr_{\bar{1}}\{R_{\bar{0}\bar{1}}(0)R_{\bar{0}\bar{1}}(2u)K_{\bar{1}}^+(u)\},\label{kcp2}
\eea where $\bar{f}(u)=-1440u(u+1)(2u+3)(2u+11)$.
Substituting Eq.(\ref{kcp2}) into (\ref{cc-2}), we obtain
\bea
\bar{t}(-u-6)=tr_{\bar{0}}\hat{T}_{\bar{0}}(u)tr_{\bar{1}}{R}_{\bar{0}\bar{1}}(0)
R_{\bar{0}\bar{1}}(2u)K^+_{\bar{1}}(u)T_{\bar{0}}(u)
V_{\bar{0}}\{K^-_{\bar{0}}(-u-6)\}^{t_{\bar{0}}}[V_{\bar{0}}^{t_{\bar{0}}}]^{-1}/\bar{f}(u).\label{cc-3}
\eea
From the Yang-Baxter equation (\ref{cc-23}) and properties (\ref{ding}), we know
\bea
&&\bar{t}(-u-6)=tr_{\bar{1}}tr_{\bar{0}}{R}_{\bar{1}\bar{0}}(0)\hat{T}_{\bar{1}}(u)
R_{\bar{0}\bar{1}}(2u)T_{\bar{0}}(u)V_{\bar{0}}\{K^-_{\bar{0}}(-u-6)\}^{t_{\bar{0}}}[V_{\bar{0}}^{t_{\bar{0}}}]^{-1}K^+_{\bar{1}}(u)/\bar{f}(u)\no\\[6pt]
&&\qquad =tr_{\bar{1}}tr_{\bar{0}}{R}_{\bar{1}\bar{0}}(0)T_{\bar{0}}(u)R_{\bar{0}\bar{1}}(2u)
\hat{T}_{\bar{1}}(u)V_{\bar{0}}\{K^-_{\bar{0}}(-u-6)\}^{t_{\bar{0}}}[V_{\bar{0}}^{t_{\bar{0}}}]^{-1}
K^+_{\bar{1}}(u)/\bar{f}(u)\no \\[6pt]
&&\qquad=tr_{\bar{1}}T_{\bar{1}}(u)tr_{\bar{0}}{R}_{\bar{0}\bar{1}}(0)
R_{\bar{0}\bar{1}}(2u)V_{\bar{0}}\{K^-_{\bar{0}}(-u-6)\}^{t_{\bar{0}}}
[V_{\bar{0}}^{t_{\bar{0}}}]^{-1}\hat{T}_{\bar{1}}(u)K^+_{\bar{1}}(u)/\bar{f}(u).\label{ding1}
\eea Substituting following identity of fused reflection matrix
\bea tr_{\bar{0}}
\{R_{\bar{0}\bar{1}}(0)R_{\bar{0}\bar{1}}(2u)V_{\bar{0}}
[K_{\bar{0}}^-(-u-6)]^{t_{\bar{0}}}[V_{\bar{0}}^{t_{\bar{0}}}]^{-1}\}=\bar{f}(u)K^-_{\bar{1}}(u).
\eea into (\ref{ding1}), we arrive at \bea
\bar{t}(-u-6)=tr_{\bar{1}}K^+_{\bar{1}}(u)T_{\bar{1}}(u)K^-_{\bar{1}}(u)\hat{T}_{\bar{1}}(u)=\bar{t}(u).\label{n-14}
\eea

\subsection*{Appendix B.3: Proof of (\ref{Crossing-3}) }

Using the crossing relation (\ref{C-S}) of the fundamental $R$-matrix and fusion technique, we can derive the corresponding relations:
\bea
&&R_{\tilde{1}2}(u)=V_{\tilde{1}}R^{t_{\tilde{1}}}_{2\tilde{1}}(-u-6)V_{\tilde{1}}^{-1},\quad
R_{2\tilde{1}}(u)=V_{\tilde{1}}^{t_{\tilde{1}}}R^{t_{\tilde{1}}}_{\tilde{1}2}(-u-6)
[V_{\tilde{1}}^{t_{\tilde{1}}}]^{-1},\no\\[4pt]
&&tr_{\tilde{1}}
\{R_{\tilde{1}\tilde{2}}(0)R_{\tilde{1}\tilde{2}}(2u)V_{\tilde{1}}
[K_{\tilde{1}}^-(-u-6)]^{t_{\tilde{1}}}[V_{\tilde{1}}^{t_{\tilde{1}}}]^{-1}\}=\tilde{f}(u)K^-_{\tilde{2}}(u),\no \\[4pt]
&&
tr_{\tilde{2}}\{R_{\tilde{1}\tilde{2}}(0)R_{\tilde{1}\tilde{2}}(2u)K_{\tilde{2}}^+(u)\}=\tilde{f}(u)V_{\tilde{1}}^{t_{\tilde{1}}}
K^+_{\tilde{1}}(-u-6)^{t_{\tilde{1}}}V_{\tilde{1}}^{-1},\label{kcp3}
\eea where $V_{ \tilde{1}}$ is a $34\times 34$ constant matrix omitted here,
$\tilde{f}(u)=-107520u(u+1)(u+2)(u+5)(2u+1)(2u+7)(2u+13)$ and
$R_{\tilde{1}\tilde{2}}(u)$ is the fused $R$-matrix with the definitions \bea
&&R_{\tilde{1}\bar{2}}(u)=[(u-1)(u+2)(u+4)(u+7)]^{-1}P^{(15)}_{23}R_{\tilde{1}2}(u+\frac12)R_{\tilde{1}3}(u-\frac12)P^{(15)}_{23},\no\\[6pt]
&&R_{\tilde{1}\tilde{2}}(u)=(u-3)^{-1}P^{(34)}_{\bar{2}3}R_{\tilde{1}\bar{2}}(u+1)R_{\tilde{1}3}(u-\frac52)P^{(34)}_{\bar{2}3}.\eea
Using the similar method as those in the previous proofs, we can show (\ref{Crossing-3}).


\section*{Appendix C: Operators product identities}
\setcounter{equation}{0}
\renewcommand{\theequation}{C.\arabic{equation}}

In order to obtain the eigenvalues of the transfer matrix $t(u)$, we need to
consider the products of two transfer matrices with shift $\delta$
of the spectral parameter \bea && t(u)t(u+\delta)=tr_a\{K_a^{
+}(u)T_a (u) K^{ -}_a(u)\hat{T}_a
(u)\}\no\\[4pt]
&&\hspace{10mm}\quad\quad\times tr_{b}\{K^{+}_{b}(u+\delta)
T_{b}(u+\delta)K^{
-}_{b}(u+\delta)\hat{T}_{b}(u+\delta)\}^{t_b}\no\\[4pt]
&&\hspace{7mm}=tr_{ab}\{K_a^{+}(u)T_a (u) K^{ -}_a(u)\hat{T}_a (u)
[T_{b}(u+\delta)K^{
-}_{b}(u+\delta)\hat{T}_{b}(u+\delta)]^{t_b}[K^{+}_{b}(u+\delta)]^{t_b}\}\no\\[4pt]
&&\hspace{7mm}=[\tilde{\rho}_{ab}(2u+\delta)]^{-1}tr_{ab} \{K_a^{
+}(u)T_a (u) K^{ -}_a(u)\hat{T}_a(u) [T_{b}(u+\delta)K^{
-}_{b}(u+\delta)\no\\[4pt]
&&\hspace{10mm}\quad\quad\times  \hat{T}_{b}(u+\delta)]^{t_b}
R_{ba}^{t_b}(2u+\delta)R_{ab}^{t_b}(-2u-2\kappa-\delta)
[K^{+}_{b}(u+\delta)]^{t_b}\}\no\\[4pt]
&&\hspace{7mm}=[\tilde{\rho}_{ab}(2u+\delta)]^{-1}tr_{ab}\{[K^{+}_{b}(u+\delta)
R_{ab}(-2u-2\kappa-\delta) K_a^{ +}(u)T_a (u) \no\\[4pt]
&&\hspace{10mm}\times  K^{ -}_a(u)\hat{T}_a
(u)]^{t_b}[R_{ba}(2u+\delta)T_{b}(u+\delta)K^{
-}_{b}(u+\delta)\hat{T}_{b}(u+\delta)]^{t_b} \}\no\\[4pt]
&&\hspace{7mm}=[\tilde{\rho}_{ab}(2u+\delta)]^{-1}tr_{ab}\{K^{+}_{b}(u+\delta)
R_{ab}(-2u-2\kappa-\delta)K_a^{ +}(u)T_a (u) \no\\[4pt]
&&\hspace{10mm}\quad\quad\times  K^{ -}_a(u)\hat{T}_a
(u)R_{ba}(2u+\delta)T_{b}(u+\delta)K^{
-}_{b}(u+\delta)\hat{T}_{b}(u+\delta) \}\no\\[4pt]
&&\hspace{7mm}=[\tilde{\rho}_{ab}(2u+\delta)]^{-1}tr_{ab}\{K^{+}_{b}(u+\delta)
R_{ab}(-2u-2\kappa-\delta) K_a^{ +}(u)T_a (u) T_{b}(u+\delta)\no\\
&&\hspace{10mm}\quad\quad\times K^{ -}_a(u)R_{ba}(2u+\delta)K^{
-}_{b}(u+\delta)
\hat{T}_a(u)\hat{T}_{b}(u+\delta)\}\no\\[4pt]
&&\hspace{7mm}=[\tilde{\rho}_{ab}(2u+\delta)]^{-1}tr_{ab}\{D_1(u)D_2(u)D_3(u)D_4(u)\},\label{FIDE}\eea
where $\kappa=6$, $\delta$ is the shift of the spectral parameter and
\bea
&&D_1(u)=K^{+}_{b}(u+\delta)R_{ab}(-2u-2\kappa-\delta) K_a^{ +}(u), \quad D_2(u)=T_a (u) T_{b}(u+\delta), \no\\[4pt]
&&D_3(u)=K^{ -}_a(u)R_{ba}(2u+\delta)K^{-}_{b}(u+\delta), \quad D_4(u)=\hat{T}_a(u)\hat{T}_{b}(u+\delta).
\eea
In the derivation, we have used the relations \bea
&&tr_{ab}\{A_{ab}^{t_a}B_{ab}^{t_a}\}=tr_{ab}\{A_{ab}^{t_b}B_{ab}^{t_b}\}
=tr_{ab}\{A_{ab}B_{ab}\},\no \\[4pt]
&& \hat{T}_a (u)  R_{ba}(2u+\delta)T_{b}(u+\delta)=
T_{b}(u+\delta)R_{ba}(2u+\delta)\hat{T}_a (u), \no \\[4pt]
 && R_{ba}^{t_b}(2u+\delta)R_{ab}^{t_b}(-2u-2\kappa-\delta)=\tilde{\rho}_{ab}(2u+\delta).\no \eea

In the definition of monodromy matrix, we introduce the inhomogeneous parameters $\{\theta_j\}$.
The role of inhomogeneous parameter and the shift $\delta$ is to generate the projectors.
Substituting $u=\theta_j$ into the term $D_2(u)$ in Eq.(\ref{FIDE}) and using the fusion relation od $R$-matrix \bea
&&R_{ai}(u)R_{bi}(u+\delta)P^{(d)
}_{ba}= P^{(d)}_{ba}R_{ai}(u)R_{bi}(u+\delta)P^{(d)}_{ba},\label{Pro-1}
\eea
we obtain
\bea &&
T_{a}(\theta_j)T_{b}(\theta_j+\delta)=R_{a1}(\theta_j-\theta_1)\cdots
R_{aj-1}(\theta_j-\theta_{j-1})R_{aj}(0)R_{aj+1}(\theta_j-\theta_{j+1})\cdots \no\\[6pt]
&&\qquad \times R_{aN}(\theta_j-\theta_{N})
R_{b1}(\theta_j-\theta_1+\delta)\cdots
R_{bj-1}(\theta_j-\theta_{j-1}+\delta)R_{bj}(\delta)\no\\[6pt]
&&\qquad \times
R_{aj}(0)R_{ja}(0)\rho_{ab}(0)^{-1}R_{bj+1}(\theta_j-\theta_{j+1}+\delta)\cdots
R_{bN}(\theta_j-\theta_{N}+\delta)\no\\[6pt]
&&=R_{jj+1}(\theta_j-\theta_{j+1})\cdots
R_{jN}(\theta_j-\theta_{N}) R_{a1}(\theta_j-\theta_1)\cdots
R_{aj-1}(\theta_j-\theta_{j-1})\no\\[6pt]
&&\qquad \times R_{b1}(\theta_j-\theta_1+\delta)\cdots
R_{bj-1}(\theta_j-\theta_{j-1}+\delta)\no\\[6pt]
&&\qquad \times P_{ba}^{(d)}S_{d} R_{ja}(0)
R_{bj+1}(\theta_j-\theta_{j+1}+\delta)\cdots
R_{bN}(\theta_j-\theta_{N}+\delta)\no\\[6pt]
&&=P_{ba}^{(d)}  R_{a1}(\theta_j-\theta_1)\cdots
R_{aj-1}(\theta_j-\theta_{j-1})R_{aj}(0)R_{ja}(0)\rho_{ab}(0)^{-1} R_{jj+1}(\theta_j-\theta_{j+1})\cdots \no\\[6pt]
&&\qquad \times R_{jN}(\theta_j-\theta_{N})
R_{b1}(\theta_j-\theta_1+\delta)\cdots
R_{bj-1}(\theta_j-\theta_{j-1}+\delta)\no\\[6pt]
&&\qquad \times R_{ba}(\delta) R_{ja}(0)
R_{bj+1}(\theta_j-\theta_{j+1}+\delta)\cdots
R_{bN}(\theta_j-\theta_{N}+\delta)\no\\[6pt]
&&=P_{ba}^{(d)}T_{a}(\theta_j)T_{b}(\theta_j+\delta).\label{FP-1}
\eea
That is
\bea
D_2(\theta_j)= P^{(d)}_{ba}D_2(\theta_j). \label{1Pro-3}
\eea
We see that the projector $P^{(d)}_{ba}$ is generated at the point of $u=\theta_j$.
From the Yang-Baxter relations (\ref{YR-1}) and the fusion relation (\ref{Pro-1}), we obtain
\bea T_a(u)T_b(u+\delta)P^{(d)}_{ba}
=P^{(d)}_{ba}T_a(u)T_b(u+\delta)P^{(d)}_{ba}.\label{Pro-5}
\eea
The projector $P^{(d)}_{ba}$ can shift from right to left in the terms in Eq.(\ref{FIDE}) and it meets the term $D_1$.
By using the fusion relation of reflection matrix
\bea
&&K^+_b(u+\delta)R_{ab}(-2u-2\kappa-\delta)K^+_a(u)P^{(d)}_{ba}\no\\[4pt]
&&\hspace{20mm}=P^{(d)
}_{ab}K^+_b(u+\delta)R_{ab}(-2u-2\kappa-\delta)K^+_a(u)P^{(d)
}_{ba},\label{Pro-4}  \eea
and the fact $P^{(d)}_{ab} =[P^{(d)}_{ab}]^2$, we obtain
\bea D_1(\theta_j)D_2(\theta_j)= D_1(\theta_j)P^{(d)}_{ba}D_2(\theta_j)= P^{(d)}_{ab}D_1(\theta_j)P^{(d)}_{ba}D_2(\theta_j)= [P^{(d)}_{ab}]^2 D_1(\theta_j)P^{(d)}_{ba}D_2(\theta_j). \label{2Pro-3} \eea
From the formula $tr_{ab}([P^{(d)}_{ab}]^2 A)=tr_{ab} (P^{(d)}_{ab}A P^{(d)}_{ab})$, we
put the projector $P^{(d)}_{ab}$ into the end in Eq.(\ref{FIDE}) and we should consider the term $D_4(\theta_j)P^{(d)}_{ab}$.
From the Yang-Baxter relation (\ref{YR-2}) and using the properties of projector, we have
\bea \hat{T}_a(u)\hat{T}_b(u+\delta)P^{(d)}_{ab}=P^{(d)
}_{ab}\hat{T}_a(u)\hat{T}_b(u+\delta)P^{(d)}_{ab},\label{Pro-6}
\eea
which gives
\bea D_4(\theta_j)P^{(d)}_{ab}= P^{(d)}_{ab}D_4(\theta_j)P^{(d)}_{ab}.\label{3Pro-3} \eea
According to the reflection equation and the properties of projector, we obtain
\bea K^-_a(u)R_{ba}(2u+\delta)K^-_b(u+\delta)P^{(d)
}_{ab}=P^{(d)
}_{ba}K^-_a(u)R_{ab}(2u+\delta)K^-_b(u+\delta)P^{(d)}_{ab},\label{Pro-3}
\eea
which gives
\bea
D_3(\theta_j)P^{(d)}_{ab}= P^{(d)}_{ba}D_3(\theta_j)P^{(d)}_{ab}.\label{4Pro-3}
\eea
Substituting Eqs.(\ref{1Pro-3}), (\ref{2Pro-3}), (\ref{3Pro-3}) and (\ref{4Pro-3}) into (\ref{FIDE}), we finally obtain \bea
&&t(\theta_j)\,t(\theta_j+\delta)=[\tilde{\rho}_{ab}(2\theta_j+\delta)]^{-1}tr_{ab}\{P^{(d)}_{ab}D_1(\theta_j)P^{(d)}_{ba}P^{(d)}_{ba}D_2(\theta_j)P^{(d)}_{ba}\no\\[6pt]
&&\hspace{5mm}\quad\quad\times
P^{(d)}_{ba}D_3(\theta_j)P^{(d)}_{ab}P^{(d)}_{ab}D_4(\theta_j)P^{(d)}_{ab}\},\label{sPro-1}\eea
which is the fusion relation.
In Eq.(\ref{sPro-1}), the $P^{(d)}_{ba}$ projects the tensor space of two auxiliary spaces ${\bf V}_a \otimes {\bf V}_b$
into the $d$-dimension fused space ${\bf V}_{\langle a b\rangle}$. Thus the product of two transfer matrices with certain spectral parameters and the fixed shift satisfy some
wonderful relations in the fused invariant subspaces.
By choosing the different values of $\delta$, we obtain different fusion relations.

It is noted that the projector can also be generated by the degenerate point of  $u=-\theta_j$ due to the fact
\bea
\hat{T}_{a}(-\theta_j)\hat{T}_{b}(-\theta_j+\delta)=P_{ab}^{(d)}\hat{T}_{a}(-\theta_j)\hat{T}_{b}(-\theta_j+\delta).\label{FP-2}\eea
Then, we shift the projector $P_{ab}^{(d)}$ by the similar ways as above and obtain
\bea
&&t(-\theta_j)t(-\theta_j+\delta)=[\tilde{\rho}_{ab}(-2\theta_j+\delta)]^{-1}tr_{ab}\{P^{(d)}_{ab}D_1(-\theta_j)P^{(d)}_{ba}P^{(d)}_{ba}D_2(-\theta_j)P^{(d)}_{ba}\no\\[6pt]
&&\hspace{5mm}\times
P^{(d)}_{ba}D_3(-\theta_j)P^{(d)}_{ab}P^{(d)}_{ab}D_4(-\theta_j)P^{(d)}_{ab}\}.\label{sPro-2}\eea

\subsection*{Appendix C.1: Proof of (\ref{Op-pro-open-2})}
When $\delta=-6$, we get an one-dimensional projector
$P_{12}^{(1)}$, whose expression is given by (\ref{pro-0}). According to the fusion of $R$-matrix, one can derive  the relations (\ref{Q-det})
which gives
\bea
&&P^{(1)}_{21}T_{1}(u)T_{2}(u-6)P^{(1)}_{21}=\prod_{i=1}^N
a(u-\theta_i)e(u-\theta_i-6)\times
{\rm id}, \no \\
&&P^{(1)}_{12}\hat{T}_1(u)\hat{T}_{2}(u-6)P^{(1)}_{12}=\prod_{i=1}^N
a(u+\theta_i)e(u+\theta_i-6)\times {\rm id}.\label{iop-0}\eea
The related fusion of reflection matrices are
\bea
&&\hspace{-2.2truecm}P^{(1)}_{21}K^-_1(u)R_{21}(2u-6)K^-_2(u-6)P^{(1)}_{12}=\no\\[4pt]
&&\hspace{0.4truecm}4(u-1)(u-6)(2u-5)(2u-1)(2u+1)\times {\rm id},\no \\[6pt]
&&\hspace{-2.2truecm}P^{(1)}_{12}K^+_2(u-6)R_{12}(-2u-2\kappa+6)K^+_1(u)P^{(1)}_{21}=\no\\[4pt]
&&\hspace{0.4truecm}-4(u+1)(u+6)(2u+5)(2u-1)(2u+1)\times {\rm id}. \label{iop-1}
\eea
Substituting equations (\ref{iop-0})-(\ref{iop-1}) into (\ref{FIDE}) and putting $u=\pm \theta_j$, $\delta=-6$, we have
\bea
&&\hspace{-1.86truecm} t(\pm\theta_j)t(\pm\theta_j-6)=4^2
\frac{(\pm\theta_j-1)(\pm\theta_j-6)
(\pm\theta_j+1)(\pm\theta_j+6)}{(\pm\theta_j-2)(\pm\theta_j-3)
(\pm\theta_j+2)(\pm\theta_j+3)}(\pm\theta_j-\frac{1}{{2}})\no\\[4pt]
&&\hspace{-1.26truecm}\times
(\pm\theta_j\hspace{-0.04truecm}-\hspace{-0.04truecm}\frac{5}{{2}})
(\pm\theta_j\hspace{-0.04truecm}+\hspace{-0.04truecm}\frac{1}{{2}})
(\pm\theta_j\hspace{-0.04truecm}+\hspace{-0.04truecm}\frac{5}{{2}})
\prod_{i=1}^N{\rho}_{12}(\pm\theta_j\hspace{-0.04truecm}-\hspace{-0.04truecm}\theta_i){\rho}_{12}
(\pm\theta_j\hspace{-0.04truecm}+\hspace{-0.04truecm}\theta_i)\hspace{-0.04truecm}\times \hspace{-0.04truecm}{\rm id},~
j=1,\cdots, N.
\label{Op-pro-open-1}
\eea
Thanks to the crossing relation of the fundamental transfer matrix $t(u)$, the above relations are equivalent to those (\ref{Op-pro-open-2}). This completes the proof of (\ref{Op-pro-open-2}).

\subsection*{Appendix C.2: Proofs of (\ref{Op-pro-open-3}) and (\ref{Op-pro-open-4})}

Substituting $u=\pm \theta_j$ and $\delta=-1$ into Eq.(\ref{FIDE}) and using
the fusion relations (\ref{1YBE15}) and (\ref{2YBE15}), we can derive the functional relations (\ref{Op-pro-open-3}).

Computing the quantity $t(u)\bar{t}(u+\delta)$ with the similar steps as Eq.(\ref{FIDE}) and substituting $u=\pm \theta_j$,
$\delta=-\frac{7}{{2}}$, $a=1$, $b=\bar1$ into the result, we can arrive at the functional relations (\ref{Op-pro-open-4}).

\subsection*{Appendix C.3: Proofs of (\ref{Op-pro-open-5}) and (\ref{Op-pro-open-6})}

The fused $R_{\tilde{1}2}(u)$ matrix (\ref{oip1}) has two
degenerate points. At the point of $u=-\frac 92$, \bea &&
R_{\tilde{1}2}(-\frac 92)= P^{(49) }_{\tilde{1}2}\times
S_{49},\label{C-34-49} \eea then we get a 49-dimensional projector
$P_{\tilde{1}2}^{(49)}$. Direct calculating gives
\bea
&&P^{(49)}_{\tilde{1}2}R_{23}(u)R_{\tilde{1}3}(u-\frac 92)P^{(49)}_{\tilde{1}2}=(u+6)S_{12}R_{13}(u-2)R_{23}(u-5)S_{12}^{-1},\no\\[6pt]
&&P^{(49)}_{2\tilde{1}}R_{32}(u)R_{3\tilde{2}}(u-\frac
92)P^{(49)}_{2\tilde{1}}
=(u+6)\tilde{S}_{12}R_{31}(u-2)R_{32}(u-5)\tilde{S}_{12}^{-1},\no\\[6pt]
&&P^{(49)}_{\tilde{1}2}K^-_2(u)R_{\tilde{1}2}(2u-\frac 92)K^-_{\tilde{1}}(u-\frac 92)P^{(49)}_{2\tilde{1}}=-2(u+1)(2u+1)(2u-1)\no\\[6pt]
&&\hspace{30mm}\times S_{12}K^-_{1}(u-2)R_{21}(2u-7)K^-_{2}(u-5)\tilde{S}_{12}^{-1},\no\\[6pt]
&&P^{(49)}_{2\tilde{1}}K^+_{\tilde{1}}(u-\frac
92)R_{2\tilde{1}}(-2u-2\kappa+\frac
92)K^+_2(u)P^{(49)}_{\tilde{1}2}
=2(u+6)(2u+5)(2u-3)\no\\[4pt]
&&\hspace{30mm}\times
\tilde{S}_{12}K^+_{2}(u-5)R_{12}(-2u-2\kappa+7)K^+_{1}(u-2)S_{12}^{-1},\label{1kcp3}
\eea where $S_{12}$ and $\tilde{S}_{12}$ are the $49\times 49$
irrelevant constant matrices. Computing the quantity $t(u)\tilde{t}(u+\delta)$ by the way as (\ref{FIDE}), substituting $u=\pm \theta_j$,
$\delta=-\frac{9}{{2}}$, $a=1$, $b=\tilde1$ in the results and using (\ref{1kcp3}),  we reach the relations (\ref{Op-pro-open-5}).

At the point of $u=-\frac {13}{2}$,
\bea &&
R_{\tilde{1}2}(-\frac {13}{2})= P^{(7) }_{\tilde{1}2}\times
S_{7},\label{C-34-7} \eea we get a 7-dimensional projector
$P_{\tilde{1}2}^{(7)}$. Direct calculating gives
\bea
&&\hspace{-1.6truecm}P^{(7)}_{\tilde{1}2}R_{23}(u)R_{\tilde{1}3}(u-\frac {13}{2})P^{(7)}_{\tilde{1}2}=(u-4)a(u)S_{1}R_{13}(u-7)S_{1}^{-1},\no\\[6pt]
&&\hspace{-1.6truecm}P^{(7)}_{2\tilde{1}}R_{32}(u)R_{3\tilde{2}}(u-\frac
{13}{2})P^{(7)}_{2\tilde{1}}
=(u-4)a(u)R_{31}(u-7),\no\\[6pt]
&&\hspace{-1.6truecm}P^{(7)}_{\tilde{1}2}K^-_2(u)R_{\tilde{1}2}(2u-\frac {13}{2})K^-_{\tilde{1}}(u-\frac {13}{2})P^{(7)}_{2\tilde{1}}\no\\[6pt]
&&\hspace{-1.2truecm}\hspace{1mm}= 4(u-4)(u-1)(2u-11)(2u-1)(2u-5)(2u-3)(2u+1)S_{1}K^-_{1}(u-7),\no\\[6pt]
&&\hspace{-1.6truecm}P^{(7)}_{2\tilde{1}}K^+_{\tilde{1}}(u-\frac
{13}{2}))R_{2\tilde{1}}(-2u-2\kappa+\frac
{13}{2})K^+_2(u)P^{(7)}_{\tilde{1}2}\no\\[6pt]
&&\hspace{-1.2truecm}\hspace{1mm}=-4(u+1)(u+6)(2u-7)(2u-5)(2u+1)(2u+3)(2u+5)
K^+_{1}(u-7)S_{1}^{-1}, \label{2kcp3} \eea where $S_{1}$ is $7\times 7$
constant matrix. Computing the quantity $t(u)\tilde{t}(u+\delta)$ by the way as (\ref{FIDE}), taking $u=\pm \theta_j$,
$\delta=-\frac{13}{{2}}$, $a=1$, $b=\tilde1$ and using the relation (\ref{2kcp3}), we arrive at the relations (\ref{Op-pro-open-6}).


\section*{Appendix D: Proofs of (\ref{sp-4}) and (\ref{sp-5})}
\setcounter{equation}{0}
\renewcommand{\theequation}{D.\arabic{equation}}

Here, we should note that the relations (\ref{sp-4})-(\ref{sp-5}) are highly non-trivial, which can be obtained as follows.
According to Eq.(\ref{oip1}), in the next fused $R_{\tilde{1}2}(u)$ matrix,
the dimension of the fused space ${\bf V}_{\tilde 1}$ is 34 and the dimension of the quantum space ${\bf V}_{2}$ is 7.
Thus the dimensional of space of $R_{\tilde{1}2}(u)$ is $34\times 7=238$.
At the point of $u=-\frac {13}{2}$, The $R_{\tilde{1}2}(u)$ reduces to a 7-dimensional projector
$P_{\tilde{1}2}^{(7)}$. In the $34\times 7-7=231$ dimensional complementary space of $P_{\tilde{1}2}^{(7)}$,
we define an operator
\bea P^{(7)\bot}_{\tilde{1}2}=1-
P^{(7)}_{\tilde{1}2}.\label{pb} \eea
Then the fusion relation (\ref{Op-pro-open-6}) can be written as
\bea
t(u)\tilde{t}(u-\frac{13}{2})=\alpha(u)t(u-7)+\gamma(u)\tilde{t}^{\bot}(u-7), \label{1Op-pro-open-6}\eea
where $\alpha(u)$ and $\gamma(u)$ are the coefficients
\bea
&&\hspace{-2.2truecm}
\alpha(u)=-2^6 \frac{(u-4)(u+1)(u+6)
}{(u-2)(u+2)
(u+3)}(u-\frac{11}{2})(u-\frac52)(u-\frac{3}{2})(u-\frac{1}{2})(u+\frac12)
(u+\frac{5}{2})\no\\[4pt]
&&\hspace{-0.26truecm}\times
\prod_{i=1}^N(u-\theta_i+4)(u+\theta_i+4)a(u-\theta_i)a(u+\theta_i), \\
&&\hspace{-2.2truecm}
\gamma(u)= \frac{u(u-\frac{11}{2})(u-\frac{3}{2}) }{2^8(u+3)
(u+2)(u+\frac32)(u+\frac12)(u-1)(u-2)(u-\frac52)(u-\frac72)}\no\\
&&\hspace{-0.26truecm}\times \prod_{i=1}^N(u-\theta_i)(u+\theta_i).
\eea
Please note $\gamma(\pm\theta_j)=0$.
Thus Eq.(\ref{1Op-pro-open-6}) reduces into (\ref{Op-pro-open-6}) naturelly. Here $\tilde{t}^{\bot}(u)$ is a new transfer matrix
with the definition
\begin{equation} \tilde{t}^{\bot}(u)= tr_{\tilde{0}^{\bot}} \{K_{\tilde{0}^{\bot}}^{   +}(u)T_{\tilde{0}^{\bot}} (u) K^{
 -}_{\tilde{0}^{\bot}}(u)\hat{T}_{\tilde{0}^{\bot}} (u)\}. \label{t-6}\end{equation}
where $\tilde{0}^{\bot}$ is a 231-dimensional auxiliary space, and
\bea
&&P^{(7)\bot}_{\tilde{1}2}K^-_2(u)R_{\tilde{1}2}(2u-\frac{13}{2})K^-_{\tilde{1}}(u-\frac{13}{2})P^{(7)\bot}_{2\tilde{1}}=u(u-\frac{11}{2})K^-_{\tilde{1}^{\bot}}(u-7),\no \\[6pt]
&&P^{(7)\bot}_{2\tilde{1}}K^+_{\tilde{1}}(u-1)R_{2\tilde{1}}(-2u-12+\frac{13}{2})K^+_2(u)P^{(7)\bot}_{\tilde{1}2}=(u-\frac32)K^+_{\tilde{1}^{\bot}}(u-7),\no \\[6pt]
&&T_{\tilde{0}^{\bot}}(u)=R_{{\tilde{0}^{\bot}}1}(u-\theta_1)R_{{\tilde{0}^{\bot}}2}(u-\theta_{2})\cdots R_{{\tilde{0}^{\bot}}N}(u-\theta_N), \no\\[6pt]
&&\hat{T}_{\tilde{0}^{\bot}} (u)=R_{N{\tilde{0}^{\bot}}}(u+\theta_N)\cdots R_{2{\tilde{0}^{\bot}}}(u+\theta_{2}) R_{1{\tilde{0}^{\bot}}}(u+\theta_1),\no \\[6pt]
&&P^{(7)\bot}_{\tilde{1}2}R_{23}(u)R_{\tilde{1}3}(u-\frac{13}{2})P^{(7)\bot}_{\tilde{1}2}=uR_{\tilde{1}^{\bot}3}(u-7),\no\\[6pt]
&&P^{(7)\bot}_{2\tilde{1}}R_{32}(u)R_{3\tilde{1}}(u-\frac{13}{2})P^{(7)\bot}_{2\tilde{1}}=uR_{3\tilde{1}^{\bot}}(u-7).
\eea
Substituting $u=0$ and $u=\frac{11}{2}$ into equation (\ref{1Op-pro-open-6}), we obtain (\ref{sp-4}) and (\ref{sp-5}), respectively.


\section*{Appendix E: Numerical solutions of (\ref{opba-1}) and (\ref{opba-2})}

\setcounter{equation}{0}
\renewcommand{\theequation}{E.\arabic{equation}}
 In this appendix, we solve the associated  BAEs (\ref{opba-1})-(\ref{x-value}) with some small sites $N$ numerically.  The results with $N=1$ and $N=2$ are shown in Table \ref{roots_real3} and \ref{roots_real}. It confirms that they  give the same
complete sets of the eigenvalues $\Lambda(u)$ by (\ref{Eigen-open-Lambda}) as those obtained by directly diagonalizing the transfer matrix (\ref{t-1}).

\begin{table}[!tbp]
\caption{Numerical solutions of the Bethe ansatz equations(\ref{opba-1})-(\ref{opba-2}), where $u=0.1\sqrt{2}, c_1=1.73$, $c_2=0.93$, $\tilde{c}_1=0.07$, $\tilde{c_2}=0.93$, $x=-2.7644$ and $N=1$.
The eigenvalues $\Lambda(u)$ calculated from the Bethe roots is exactly the same as that from the exact diagonalization of the transfer matrix (\ref{t-1}).
We note that the dimension of Hilbert space is $7$ and there are only 4 energy levels due to the degeneracy of eigenvalues.  }\label{roots_real3}
\vspace{0.38truecm}
{
\begin{tabular}{|c|c|c|c|c|c|}
\hline $ u_1^{(1)} $ & $ u_2^{(1)}  $ &$ u_1^{(2)} $ & $  u_2^{(2)}$ & $ \Lambda(u) $ & $ n $\\ \hline

$--$&$--$&$--$&$--$&-5951.5545 & 1 \\ \hline

-0.935732084981 &$--$& -0.708786384923 & $--$  & -4691.3338  & 2\\ \hline

-0.267170490371 &$--$&  -1.64756781842i & $--$ & -2427.0809 & 3\\ \hline

2.09276820469i &-0.119459001451i &  -1.74925136686i & -2.67861768166i  & -1906.599 & 4\\ \hline

\end{tabular}
}
 \end{table}

\begin{table}[!ht]
\caption{Numerical solutions of the Bethe ansatz equations (\ref{opba-1})-(\ref{opba-2}), where $u=0.1\sqrt{2}, c_1=1.73$, $c_2=0.93$, $\tilde{c}_1=0.07$, $\tilde{c_2}=0.93$, $x=-2.7644$ and $N=2$.
Here $n$ indicates the number of the energy levels and $\Lambda(u)$ is the eigenvalues of transfer matrix (\ref{t-1}).
The eigenvalues $\Lambda(u)$ calculated from the Bethe roots is exactly the same as that from the exact diagonalization of the transfer matrix. }\label{roots_real}
\vspace{0.38truecm}
{
\footnotesize
\begin{tabular}{|c|c|c|c|c|c|}
\hline $ u_1^{(1)} \quad u_2^{(1)} $ & $ u_3^{(1)} \quad u_4^{(1)} $ &$ u_1^{(2)} \quad u_2^{(2)} $ & $  u_3^{(2)} \quad u_4^{(2)}$ & $ \Lambda(u) $ & $ n $\\ \hline

$--$ & $--$ & $--$ & $--$  & $-5016141.23085$ & $1$ \\ \hline

 $\makecell{-3.41283959954\\-1.85092085347i}$ & $-3.81975270035 $ &
 $\makecell{-4.14735396445\\+1.76207925802i}$ & $\makecell{4.14735396445\\+1.76207925802i}$ & $ \multirow{2}{*}{-4821096.6438} $  & \multirow{2}{*}{2}
 \\ \cline{1-4}
 $\makecell{-3.41283959954\\+1.85092085347i}$ & $--$ & $-3.09039102921i$ & $--$ &  &  \\ \hline

 $\makecell{1.65848775056\\+0.967805933636i}$ & $\makecell{1.65848775056\\-0.967805933636i}$ & $
\makecell{2.2666035361\\+1.91447835223i}  $ & $  \makecell{-2.2666035361\\+1.91447835223i} $ & \multirow{2}{*}{-4729402.04678}  & \multirow{2}{*}{3} \\ \cline{1-4}
$6.7026965576i$  & $-4.37058021448i$ & $
\makecell{-0.354985269568\\+6.89218164042i} $ & $ \makecell{-0.354985269568\\-6.89218164042i}$ & &\\ \hline

$1.99667248616$ & $--$ & $-2.39612116572$ & $--$ & $-4694200.0595$ & $4$\\ \hline

$\makecell{1.68137688114\\+1.05560185579i}$ & $ \makecell{1.68137688114\\-1.05560185579i}  $ & $ \makecell{2.24826247491\\-2.03965964713i} $ & $  \makecell{-2.24826247491\\-2.03965964713i}$ & $-4688487.84325$ & $5$\\ \hline

$1.26204946337 $ & $   \makecell{-1.14724900956\\+1.16711191966i} $ & $ 0.545790518363  $ & $ -4.69763384402i $
 & \multirow{2}{*}{-4233552.12081} & \multirow{2}{*}{6}\\ \cline{1-4}
  $\makecell{1.14724900956\\+1.16711191966i} $ & $--$ & $-1.99229320667i$ & $--$  &  &  \\ \hline

  $1.55535359162i$ & $ -3.2211592858i$ & $  -1.24093466186i $ & $   -1.97376155171i $ &
 \multirow{2}{*}{-3975340.70306}  & \multirow{2}{*}{7} \\ \cline{1-4}
$\makecell{0.536196472927\\-0.52280662971i}$ & $\makecell{-0.536196472927\\-0.52280662971i}$ & $
 4.39556025287i $ & $ -6.07697315709i$ &   & \\ \hline

  $2.8830549976i$ & $\makecell{0.438607392786\\-0.514898897408i}$ & $
\makecell{0.194603453959\\-1.64190120059i}  $ & $  \makecell{ 0.194603453959\\+1.64190120059i} $ &
\multirow{2}{*}{-3907008.17938}  & \multirow{2}{*}{8} \\ \cline{1-4}
$\makecell{ -0.438607392786\\-0.514898897408i}$ & $ 1.48607189256i$ & $
\makecell{ 0.38129921316\\-5.41380193143i}  $ & $  \makecell{-0.38129921316\\-5.41380193143i} $ &   & \\ \hline

  $0.834074424847$ & $--$ & $1.79644377488i$ & $--$  & $-3734114.09685$ & $9$ \\ \hline

$\makecell{ 2.17340767734\\-2.06923337856i}$ & $ 0.793074552023 $ & $
\makecell{2.94598356295\\+1.80122292837i}  $ & $-5.91214285092i $ &
\multirow{2}{*}{-3478542.02078}  & \multirow{2}{*}{10} \\ \cline{1-4}
$ 2.48907525716$ & $\makecell{-2.17340767734\\-2.06923337856i}$ & $
\makecell{ 2.94598356295\\-1.80122292837i}  $ & $  -2.02880536484i $&& \\ \hline

$0.500000344876i$ & $-0.499999655124i$ & $2.17455120807i$ & $0.76676387982i$  & $-2858749.59524$ & $11$ \\ \hline

$4.58526327893i$ & $  -6.74194055084i$ & $
0.961258372949i$ & $   -6.32116950936i $ & \multirow{2}{*}{-2856126.22781}  & \multirow{2}{*}{12} \\ \cline{1-4}
$ -0.203860238421i$ & $ -0.785282106992i$ & $
2.13965830347i  $ & $-7.53145618958i$&   & \\ \hline

$0.499998640486i$ & $-0.500001359514i$ & $ -3.94121010562i$ & $-1.47412161487i$  & $-2811189.7622$ & $13$ \\ \hline

  $-3.35503145999i$ & $-0.140014243711$ & $ -4.80211808166i $ & $ -1.11991673392i $ & \multirow{2}{*}{-2763676.66836} & \multirow{2}{*}{14} \\ \cline{1-4}
  $ -0.955849571444i $ & $--$ & $-2.16694817984i$ &$--$ &  &  \\ \hline

  $0.714483559325i $ & $  0.28993729611i  $ & $   -3.96807488341i  $ & $  -1.52177099594i $ & $-2653703.59277$ & $15$ \\ \hline

-1.94886309675i & 0.00854346044972 &  1.4951752689i &
 -5.12293053373i& \multirow{2}{*}{-2527008.4076} & \multirow{2}{*}{16} \\ \cline{1-4}
$-0.996068427755i$ &  $--$ &   $-0.339105352313$ &$--$  & & \\ \hline

-2.04233688822i&-0.994403502778i &  1.49335108691i  &
 -5.13002381492i& \multirow{2}{*}{-2512168.25011} & \multirow{2}{*}{17} \\ \cline{1-4}
$0.00810826885271$ & $--$  &   $-0.798668856439i$ & $--$& & \\ \hline

-0.299733444106 & $--$ &  -1.24859483198i&  $--$ & $-2185886.82673$  & $18$\\ \hline

$\makecell{2.52810982948\\-1.77298733443i}$ & $-0.325136437254$ & $
\makecell{3.13415675936\\+1.6618407843i}  $ & $ -5.6853558594i $ & \multirow{2}{*}{-2153047.48175}  & \multirow{2}{*}{19} \\ \cline{1-4}
$ 2.88694568567$ & $\makecell{ -2.52810982948\\-1.77298733443i}$ & $
\makecell{ 3.13415675936\\-1.6618407843i}  $ & $ 1.17077143475i$ &   & \\ \hline

\end{tabular}
}
 \end{table}

 \begin{table}[!ht]
{
\footnotesize
\begin{tabular}{|c|c|c|c|c|c|}
\hline

3.44398394461i&-0.0274565197948 &  1.6008364445i  &
 -2.03897325162i& \multirow{2}{*}{-1969244.31112} & \multirow{2}{*}{20} \\ \cline{1-4}
$-1.29987327316i$ &  $--$  &   $4.93464039124i$ & $--$ & & \\ \hline

$-4.81562024359i$ & $6.94613844262i$ & $
 -2.02473066893i  $ & $   -7.62002595284i$ & \multirow{2}{*}{-1920491.19946}  & \multirow{2}{*}{21} \\ \cline{1-4}
$0.0251756453507i$ & $-1.33268061287i$ & $
1.60954720215i  $ & $ -6.62520887432i $ &   & \\ \hline

-0.0323238438568i & 1.36714006951i & -1.61177416733i &   -2.08522100858i & $-1861114.04676$  & $22$\\ \hline

-0.125208316202 & $--$ & -1.57653141173i&  $--$ & $-1693829.05475$  & $23$\\ \hline

$\makecell{2.38283514748\\-1.85909026247i}$ & $-0.110410785338$ & $
\makecell{3.06630502088\\+1.70701823797i} $ &  -5.78331662181i &\multirow{2}{*}{ -1581425.06425}  & \multirow{2}{*}{24} \\ \cline{1-4}
$ 2.75499163166$ & $\makecell{-2.38283514748\\-1.85909026247i}$ & $
\makecell{3.06630502088\\-1.70701823797i}  $ & $   -1.59055942502i$ &   & \\ \hline

-0.0330168898996 & -1.09302022855 & 1.60778023783i &   -0.617221191596i & $-1300823.23712$  & $25$\\ \hline

$-0.024378362921$ & $6.52019225419i$ & $
 -1.60582926459i  $ & $  -6.0376219633i $ & \multirow{2}{*}{-1297366.8296}  & \multirow{2}{*}{26} \\ \cline{1-4}
$1.02231286722$ & $ -4.3123285693i$ & $
7.38444875229i  $ & $  -0.83571853053i$ &   & \\ \hline

-0.0234287328835i &3.05685393453i &  -1.60604263198i   &
-4.55223859133i& \multirow{2}{*}{-1269072.06287} & \multirow{2}{*}{27} \\ \cline{1-4}
$-0.953384128998$ & $--$  &   $-0.951046954716i$ &$--$ & & \\ \hline

$2.67599832211i$ & $ 0.49651278434$ & $
1.53514204234i$ & $  -1.41650776076i$  & \multirow{2}{*}{-1103630.738}  & \multirow{2}{*}{28} \\ \cline{1-4}
$0.0458065331543$ & $1.4304772125i$ & $
\makecell{-0.385648315633\\-5.36261136433i}  $ & $  \makecell{0.385648315633\\-5.36261136433i} $ &   & \\ \hline

$0.519228459967$ & $  -3.03957626009i$ & $
\makecell{-0.0592761947403\\-1.51072452207i}  $ & $  \makecell{0.0592761947403\\-1.51072452207i} $ & \multirow{2}{*}{-1073809.82546}  & \multirow{2}{*}{29} \\ \cline{1-4}
$0.0333881032309$ & $ -1.61436064678i$ & $
 4.55797646369i  $ & $ -5.98458735784i  $ &   & \\ \hline

 -1.87712225512i  & 0.0275778571429i  & \makecell{-0.0653873440111\\-1.52113713368i}   &
\makecell{-0.0653873440111\\+1.52113713368i} & \multirow{2}{*}{-1057307.42088} & \multirow{2}{*}{30} \\ \cline{1-4}
$-0.577725269181$ &  $--$ &   $-4.97321493303i$ & $--$ & & \\ \hline

\end{tabular}
}
 \end{table}

\newpage


\begin{thebibliography}{99}
\bibitem{1} R. J. Baxter, {\it Exactly Solved Models in Statistical Mechanics},
Academic Press, 1982.
\bibitem{2} C. N. Yang, {\it Some exact results for the many-body problem in one dimension with repulsive
delta-function interaction, Phys. Rev. Lett.} {\bf 19} (1967) 1312.
\bibitem{3} V.\,E. Korepin, N.\,M. Bogoliubov and A.\,G. Izergin,
{\it Quantum Inverse Scattering Method and Correlation
Function}, Cambridge University Press, 1993.

\bibitem{4} T. Giamarchi, {\it Quantum Physics in One Dimension}, Oxford University Press, Oxford, 2003.

\bibitem{A-1} H.J. de Vega, A. Gonzalez-Ruiz,  {\it Exact solution of the
$SU_q(n)$ invariant quantum spin chains,} {\it Nucl. Phys.} {\bf B
417} (1994) 553.

\bibitem{A-2} W.L. Yang and Y.Z. Zhang, {\it Exact solution of the $ A^{ (1)}_{
n-1}$ trigonometric vertex model with non-diagonal open
boundaries}, {\it JHEP} {\bf 01} (2005) 021 [hep-th/0411190].

\bibitem{A-3}  J. Cao,  W.-L. Yang, K. Shi, Y. Wang, {\it  Nested off-diagonal
Bethe ansatz and exact solutions of the $SU(N)$ spin chain with
generic integrable boundaries,} {\it JHEP} {\bf 04} (2014) 143.


\bibitem{B-1} N.Yu. Reshetikhin, {\it The Spectrum of the transfer matrices
connected with Kac-Moody algebras}, {\it Lett. Math. Phys.} {\bf
14} (1987) 235.
\bibitem{B-2} M. J. Martins and P. B. Ramos, {\it The algebraic Bethe ansatz for rational braid-monoid lattice
models}, {\it Nucl. Phys.} {\bf B 500} (1997) 579.
\bibitem{B-3} S. Artz, L. Mezincescu and R.I. Nepomechie, {\it Analytical Bethe
ansatz for $A^{ (2)}_{ 2n-1}$, $B^{ (1)}_n$, $C^{ (1)}_n$, $D^{(1)}_n$ quantum algebra invariant open spin chains}, {\it J. Phys.} {\bf A 28}  (1995) 5131 [hep-th/9504085].

\bibitem{B-4} G. -L. Li, K. J. Shi and R. H. Yue, {\it Nested Bethe ansatz for the $B_N$ vertex model with open boundary conditions,} {\it Nucl.
Phys.} {\bf B 696} (2004) 381.



\bibitem{B-5}  G.-L. Li,  K.J. Shi,{\it The algebraic Bethe ansatz for open vertex models}, {\it J. Stat. Mech.}  {\bf 25}  (2007) P01018.


\bibitem{C-1} G. -L. Li, K. J. Shi and R. H. Yue, {\it Algebraic Bethe Ansatz Solution to $C_N$ Vertex Model with Open Boundary Conditions}, {\it Commun. Theor. Phys.} {\bf 44} (2005) 89.

\bibitem{C-2} D. Chicherin, S. Derkachov and A. P. Isaev, {\it The spinorial R-matrix,} {\it J. Phys.} {\bf A 46} (2013) 485201.
\bibitem{C-3} G. A. P. Ribeiro, {\it On the partition function of the $Sp(2n)$ integrable vertex model,} {\it J. Stat. Mech.} (2023) 043102 arXiv:2211.06487.

\bibitem{5} F. C. Alcaraz, M. N. Barber, M. T. Batchelor, R. J. Baxter and G. R. W. Quispel, {\it Surface exponents of
the quantum XXZ, Ashkin-Teller and Potts models, J. Phys.} {\bf A 20} (1987) 6397.

\bibitem{i1} S. T. Carr and A.M. Tsvelik, {\it Spectrum and correlation functions of a quasi-one dimensional quantum Ising model}, {\it Phys. Rev. Lett.} {\bf 90} (2003) 177206.
\bibitem{i2} R. Coldea, D. A. Tennant, E.M. Wheeler, E. Wawrzynska, D. Prabhakaran, M. Telling, K. Habicht, P. Smibidl, and K. Kiefer, {\it Quantum criticality in an Ising chain: experi
mental evidence for emergent $E_8$ symmetry}, {\it Science} {\bf 327} (2010) 177180.
\bibitem{i3} D. Borthwick and S. Garibaldi, {\it Did a 1-dimensional magnet detect a 248-dimensional Lie algebra?}, {\it Not. Amer. Math. Soc.} {\bf 58} (2011) 1055.



\bibitem{Li23}C. Li, V. L. Quito, D. Schuricht and P. L. S. Lopes, {\it $G_2$ integrable point characterization via isotropic spin-3 chains}, {\it Phys. Rev.} {\bf B 108} (2023) 165123.
\bibitem{Hu18} Y. Hu and C. L. Kane, {\it Fibonacci topological superconductor}, {\it Phys. Rev. Lett.} {\bf 120} (2018) 066801.
\bibitem{Lop19} P. L. S. Lopes, V. L. Quito, B. Han and J. C. Y. Teo, {\it Non-Abelian twist to integer quantum Hall states}, {\it Phys. Rev.} {\bf B 100} (2019) 085116.
\bibitem{Nay08} C. Nayak, S. H. Simon, A. Stern, M. Freedman and S. Das Sarma, {\it Non-Abelian anyons and topological quantum computation}, {\it Rev. Mod. Phys.} {\bf 80} (2008) 1083.




\bibitem{john1} J. C. Baez, {\it The octonions}, {\it American Mathematical Society} {\bf 39} (2001) 145.
\bibitem{john2} D. D. Joyce, {\it Compact Riemannian 7-manifolds with holonomy $G_2$}, {\it J. Differential Geometry} {\bf 43} (1996) 291.
\bibitem{john3} S. Karigiannis, N. C. Leung and J. D. Lotay, {\it Lectures and Surveys on $G_2$-Manifolds and Related Topics}, Springer Press, (2020).

\bibitem{Ogi86} E. I. Ogievetsky, {\it Factorized S-matrix with $G_2$ symmetry}, {\it J. Phys.} {\bf G 12} (1986) L105.
\bibitem{Mac91} N. J. MacKay, {\it Rational R-matrices in irreduciable representations}, {\it J. Phys.} {\bf A 24} (1991) 4017.

\bibitem{martins} M. J. Martins,{\it The spectrum properties of an integrable $G_2$ invariant vertex model},{\it Nucl. Phys.} {\bf B 989} (2023) 116131.

\bibitem{yung} C. M. Yung, M. T. Batchelor, {\it Diagonal $K$-matrices and transfer matrix eigenspectra associated with the $G^{(1)}_2$ R-matrix}, {\it Phys. Lett.} {\bf A 198} (1995) 395.


\bibitem{re-1} E.K. Sklyanin, {\it Boundary Conditions for Integrable Quantum
Systems}, {\it J. Phys. A} {\bf 21} (1988) 2375.

\bibitem{re-2} L. Mezincescue and R.I. Nepomechie, {\it Integrable open spin
chains with nonsymmetrical R-matrices}, {\it J. Phys. A} {\bf 24}
(1991) L17.
\bibitem{a1} L. A. Takhtadzhan and L. D. Faddeev, {\it The quantum method of the inverse problem and the Heisenberg
XYZ model, Rush. Math. Surveys} {\bf 34} (1979) 11.

\bibitem{a2}  N. Yu. Reshetikhin, {\it The functional equation method in the theory of exactly soluble quantum
systems, Sov. Phys. JETP} {\bf 57} (1983) 691.

\bibitem{f1} M. Karowski, {\it On the bound state problem in $1+1$ dimensional field theories, Nucl. Phys.} {\bf B 153} (1979) 244.
\bibitem{f2} P. P. Kulish, N. Yu. Reshetikhin and E. K. Sklyanin, {\it Yang-Baxter equation and representation theory:
I, Lett. Math. Phys.} {\bf 5} (1981) 393.
\bibitem{f3} P. P. Kulish and E. K. Sklyanin, {\it Quantum spectral transform method recent developments, Lecture Notes
in Physics,} {\bf 151} (1982) 61.
\bibitem{f4} A. N. Kirillov and N. Yu. Reshetikhin, {\it Exact solution of the Heisenberg XXZ model of spin $s$, J. Sov. Math.} {\bf 35} (1986) 2627.
\bibitem{f5} A. N. Kirillov and N. Yu. Reshetikhin, {\it Exact solution of the integrable XXZ Heisenberg model with
arbitrary spin I. The ground state and the excitation spectrum, J. Phys.} {\bf A 20} (1987) 1565.
\bibitem{f6} L. Mezincescu and R. I. Nepomechie, {\it Fusion procedure for open chains, J. Phys.} {\bf A 25} (1992) 2533.
\bibitem{f7} L. Mezincescu and R. I. Nepomechie, {\it Analytical Bethe Ansatz for quantum algebra invariant spin
chains, Nucl. Phys.} {\bf B 372} (1992) 597.




\bibitem{Cao1} J. Cao, W.-L. Yang, K. Shi and Y. Wang, {\it Off-diagonal Bethe Ansatz and exact solution of a
topological spin ring, Phys. Rev. Lett.} {\bf 111} (2013) 137201.

\bibitem{Cao13} J. Cao, W.\,-L. Yang, K. Shi and Y. Wang, {\it Off-diagonal Bethe ansatz solutions of the anisotropic
spin-$\frac12$ chains with arbitrary boundary fields}, {\it Nucl. Phys.} {\bf B 877} (2013) 152.

\bibitem{Cao14} J. Cao, W.\,-L. Yang, K. Shi and Y. Wang, {\it Nested off-diagonal Bethe ansatz and exact solutions of the
$su(n)$ spin chain with generic integrable boundaries}, {\it JHEP} {\bf 04} (2014) 143.

\bibitem{Hao14} K. Hao, J. Cao, G.-L. Li, W.-L. Yang, K. Shi and Y. Wang,
Exact solution  of the Izergin-Korepin model with general non-diagonal boundary terms,
{\it JHEP} {\bf 06} (2014) 128.



\bibitem{Cao2} Y. Wang, W. -L. Yang, J. Cao and K. Shi, {\it Off-Diagonal Bethe
Ansatz for Exactly Solvable Models}, Springer Press, 2015.


\end{thebibliography}
\end{document}